\documentclass[pre,aps,twocolumn,showpacs]{revtex4}
\usepackage{graphicx}

\begin{document}

\title{ Electronic specific heat of DNA: Effects of backbones and disorder }

\author{Sourav Kundu}

\email{sourav.kunduphy@gmail.com}

\affiliation{Condensed Matter Physics Division, 
Saha Institute of Nuclear Physics, 
1/AF, Bidhannagar, Kolkata 700 064, India}

\author{S. N. Karmakar}

\affiliation{Condensed Matter Physics Division, 
Saha Institute of Nuclear Physics, 
1/AF, Bidhannagar, Kolkata 700 064, India}

\begin{abstract}
In this present work we report the results of our investigation 
on the electronic specific heat (ESH) of DNA molecule modelled 
within the tight-binding framework. We take four different DNA 
sequences ranging from periodic, quasi-periodic to random and 
studied both ESH and also the density of states to supplement 
our ESH results. The role of the backbone structure and the effects 
of environment on ESH are discussed. We observe that irrespective 
of the sequences there is a universal response of the ESH spectra 
for a given disorder. The nature of response of specific heat on 
backbone disorder is totally opposite in low and high temperature 
regimes.

\end{abstract}
  

\maketitle

\section{Introduction}
Since the last decade, interest in studying the electrical properties of 
DNA has enhanced considerably as it appears that DNA can conduct electrons, 
and becomes a promising agent for future nanoelectronic devices which can 
help to overcome the drawbacks like energy efficiency of present silicon-based 
devices. Not only this, it has many advantageous properties like self-assemble 
and self-replication which can make it possible to produce nanostructures 
with greater precision that is not achievable with classical silicon-based 
technology~\cite{winfree}. The question whether DNA or in broad aspect biomolecules 
can conduct electrons or not, was first addressed by Eley and Spivey~\cite{eley} 
in 1962. The first effort to use organic molecules as electronic components 
started also quite early in 1974~\cite{aviram}, but the lack of knowledge 
of electrical properties of DNA made the task quite difficult. With time, 
the advent of new-generation sophisticated techniques and low temperature 
measurement facilities make it possible to investigate the physical properties 
of DNA and other biomolecules. Till date a large number of studies have been 
performed on electronic transport properties of DNA. A various number of 
different techniques have been applied which include measurements on single-stranded 
DNA~\cite{braun, porath}, measurements on oriented DNA strands on lipid films~\cite{okahata}, 
measurement on the current-voltage (I-V) characteristics of unordered DNA on 
the nanocontacts~\cite{lee, otsuka}, and also on artificially made poly(GC) 
chains~\cite{fink}. Despite these vast efforts no conclusive results appear, 
on contrary DNA behaves in various modes in electrical conduction, such 
as an insulator~\cite{storm}, wide-band gap semiconductor~\cite{porath}, 
ohmic~\cite{fink}, and even proximity-induced superconductor at low 
temperature~\cite{kasumov}. This large variety of experimental results 
led to various theoretical models in which electrical transport is mediated  
by polarons~\cite{conwell}, solitons~\cite{hermon}, electrons and 
holes~\cite{dekker, ratner, beratan}. Thus electronic transport properties 
of DNA still remain quite debatable and Ref.~\cite{endres, alburev} provide 
a vivid review. 

While there is so much effort to understand the electrical properties of DNA, 
only mere number of attempts have been made to study its thermal and thermodynamic 
properties~\cite{moreira1, sarmento, moreira2, moreira3}. Recently a study~\cite{mendes} 
showed that the knowledge of thermal properties of different biomolecules (e.g., 
polypeptides etc. ) may become helpful in the determination process of different 
neurodegenerative diseases like Alzheimer and Parkinson. In our work we would like 
to extend the knowledge of thermal properties of DNA through study of specific heat 
in details. Our main aim is to investigate the role of backbones on the 
electronic specific heat (ESH) spectra, with the purpose of finding the 
likeness and dis-likeness among them and to predict some kind of standard 
behavior. We model DNA within the tight-binding framework, incorporating its 
backbone structure which was lacking in the previous studies~\cite{moreira1, 
sarmento, moreira2, moreira3, mauriz, albu}, to investigate the effects of 
the presence of backbones and also of the environment. We use four DNA sequences 
ranging from periodic, quasi-periodic to random, and see that irrespective of 
sequences the features of specific heat are similar in the clean case (no disorder) 
and this behaviour prevails even in presence of environmental effects at any 
particular disorder. 

The rest of the paper is organized in the following way: In Sec. 2 
we present the tight-binding Hamiltonian and discuss about our 
theoretical formulation. In Sec. 3 we show our numerical results 
and finally summarize in Sec. 4.

\section{Model and Theoretical Formulation}
To study the ESH spectra of DNA molecule we use the so-called dangling 
backbone ladder model (DBLM)~\cite{klotsa, gcuni} within the tight-binding 
framework, as it is more realistic than the simple ladder model 
since the former one incorporates the backbone structure of DNA. As the sugar-phosphate 
backbones are negatively charged and form the outer part of the double-helix, they 
can easily interact with environment and the substrates on which experiments are 
performed and the backbone site energies get changed randomly~\cite{storm, kasumov, 
barnett, tran, zhang, pablo, lcai, sourav}. Therefore by introducing disorder in the 
backbone site-energies we can incorporate environmental effects into the system 
which quite well replicate the experimental situations.

\begin{figure}[ht]
\centering

    \includegraphics[width=60mm]{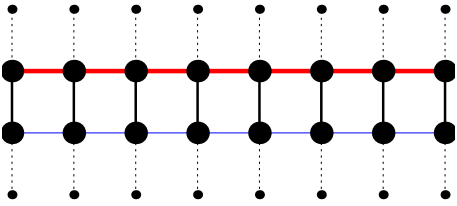}

\caption{(Color online) Pictorial view of the dangling backbone ladder 
model to study ESH spectra of double-stranded DNA. The red (thick) and 
blue (thin) lines are the two strands and the large solid dots on them 
are the nucleotides. The small black dots on the upper and lower sides 
of the ladder represent the backbone sites and the dotted lines between 
the two types of dots represent the coupling between the nucleotides with 
the corresponding backbone sites.}

\label{fig1}
\end{figure}

The effect of environment on the thermal properties of DNA, specially on 
the specific heat is so far not well-explored, hence for this purpose we 
use the following tight-binding Hamiltonian of the DBLM to mimic DNA molecule 
(see Fig.~\ref{fig1}) 

\begin{equation}
 H_{DNA} = H_{ladder}+ H_{backbone}~,
\label{hamilton}
\end{equation}

\noindent
where, 
\begin{eqnarray}
& H_{ladder}&= \sum\limits_{i=1}^N\sum\limits_{j=I,II}\left(\epsilon_{ij}
c^\dagger_{ij}c_{ij}
+t_{ij}c^\dagger_{ij}c_{i+1j}+\mbox{H.c.} \right)\nonumber \\
&&~~~~~~~~~~~~~~~~~+ \sum_{i=1}^N v \left(c^\dagger_{iI}c_{i II}+
\mbox{H.c.} \right)~, \\
& H_{backbone}&=\sum\limits_{i=1}^N\sum\limits_{j=I,II}
\left(\epsilon_i^{q(j)}c^\dagger_{i q(j)}c_{i q(j)}\right.\nonumber \\
&&~~~~~~~~~~~~~~~~~+\left.t_i^{q(j)}c^\dagger_{ij}c_{i q(j)}+
\mbox{H.c.} \right)~,
\end{eqnarray} 
where $c_{ij}^\dagger$ and $c_{ij}$ are the electron creation and annihilation 
operators at the {\it i}th nucleotide at the jth stand, $t_{ij}=$ nearest 
neighbour hopping amplitude between nucleotides along the jth branch of the 
ladder, $\epsilon_{ij}=$ on-site energy of the nucleotides, $\epsilon_{i}^{q(j)}=$ 
on-site energy of the backbone site adjacent to ith nucleotide of the jth 
strand with $q(j)=\uparrow,\downarrow$ representing the upper and lower 
strands respectively, $t_{i}^{q(j)}=$ hopping amplitude between a nucleotide 
and the corresponding backbone site, and $v=$ vertical hopping between 
nucleotides in the two strands of the ladder. For simplicity, we set 
$\epsilon_i^{q(j)}=\epsilon_b$, $t_{ij}=t_i$ and $t_i^{q(j)}=t_b$. 

To study the electronic specific heat of the DNA molecule, we use the 
most general formalism where the specific heat at constant volume is 
given by the partial derivative of average energy of the system 
with respect to temperature 
\begin{equation}
 C_v = \frac{\partial <E>}{\partial T}
\end{equation}
where, 
\begin{eqnarray}
& <E>& = \sum\limits_{i=1}^N (E_i-\mu) f(E_i) \\
& f(E_i)&= \frac {1} {1+\exp(\frac{E_i-\mu}{k_BT})}
\end{eqnarray} 
where $<E>$ is the average energy of the system, $E_i$ the energy of an 
electron at the {\it i}th eigenstate, $\mu$ is the chemical potential, 
T is the temperature, $k_B$ is the Boltzmann constant and $f(E_i)$ is 
the occupation probability of the ith eigenstate according to Fermi-Dirac 
statistics. Using the expressions of $<E>$ and $f(E_i)$ we find the 
following expression for electronic specific heat (ESH) of DNA 
\begin{equation}
 C_v = \sum\limits_{i=1}^N\frac{(E_i-\mu)^2 \exp(\frac{E_i-\mu}{k_B T})}{{k_B}T^2 (\exp(\frac{E_i-\mu}{k_B T})+1)^2}
\end{equation} 

To explain the ESH spectra we also study the electronic 
density of states (DOS) of the system. We use Green's 
function formalism to find DOS of the system which is 
given by 
\begin{equation}
 \rho(E) = - \frac{1}{\pi} {\rm Im[Tr[G(E)]]}
\end{equation}
where, 
$G(E)= (E-H+i\eta)^{-1}$ is the Green's function for the entire DNA molecule 
with electron energy E as $\eta\rightarrow0^+$, $H=$ Hamiltonian of the 
DNA, and, ${\rm Im}$ and ${\rm Tr}$ respectively represents imaginary 
part and trace over the entire Hilbert space. 

\section{Results and Discussions}
To study the ESH of DNA, we take four different DNA sequences, 
of which two are periodic, one is random and another is quasi-periodic. 
In the present work we take Fibonacci sequence as a prototype example 
of a quasiperiodic sequence which is derived using the inflation rule : 
A$\rightarrow$AT, T$\rightarrow$A. In order to investigate the effect 
of backbone environment, we first study the simple ladder model without 
any backbone and then the  dangling backbone ladder model, which incorporates 
the backbone structure. In DBLM environmental fluctuations are incorporated 
introducing disorder into the backbone sites. To represent the actual experimental 
situation, in this work we simulate the environmental fluctuations including also 
the effect of water environment by distributing the backbone site-energy $\epsilon_b$ 
randomly within the range $[\bar\epsilon_b$-w$/2,\bar\epsilon_b$+w$/2]$, where 
$\bar\epsilon_b$ represents the average backbone site energy and w being the 
disorder strength. To have a physical insight about the ESH of DNA we have 
also evaluated the DOS of the four DNA sequences as stated earlier for various 
disorder strengths (w). For numerical calculations the on-site energies of the 
nucleotides ($\epsilon_{ij}$) are taken as the ionization potential and the 
following numerical values are used in our work: $\epsilon_G$= 8.177 eV, 
$\epsilon_C$= 9.722 eV, $\epsilon_A$= 8.631 eV, $\epsilon_T$= 9.464 eV. The 
intrastrand hopping amplitude between identical neighbouring bases is taken 
as t= 0.35 eV while that between unlike nucleotides is taken as t= 0.17 eV. 
We take interstrand hopping parameter {\it i.e.}, vertical hopping to be v= 0.3 eV. 
The parameters used here are adopted from the first-principle calculations~\cite{voit, yan, senth}. 
The hopping between a nucleotide and corresponding backbone is taken as 
$t_b$= 0.7 eV~\cite{sourav, cuni}. We also set $k_B$=1. 

\begin{figure}[ht]
  \centering
  \begin{tabular}{c}
    \includegraphics[width=60mm]{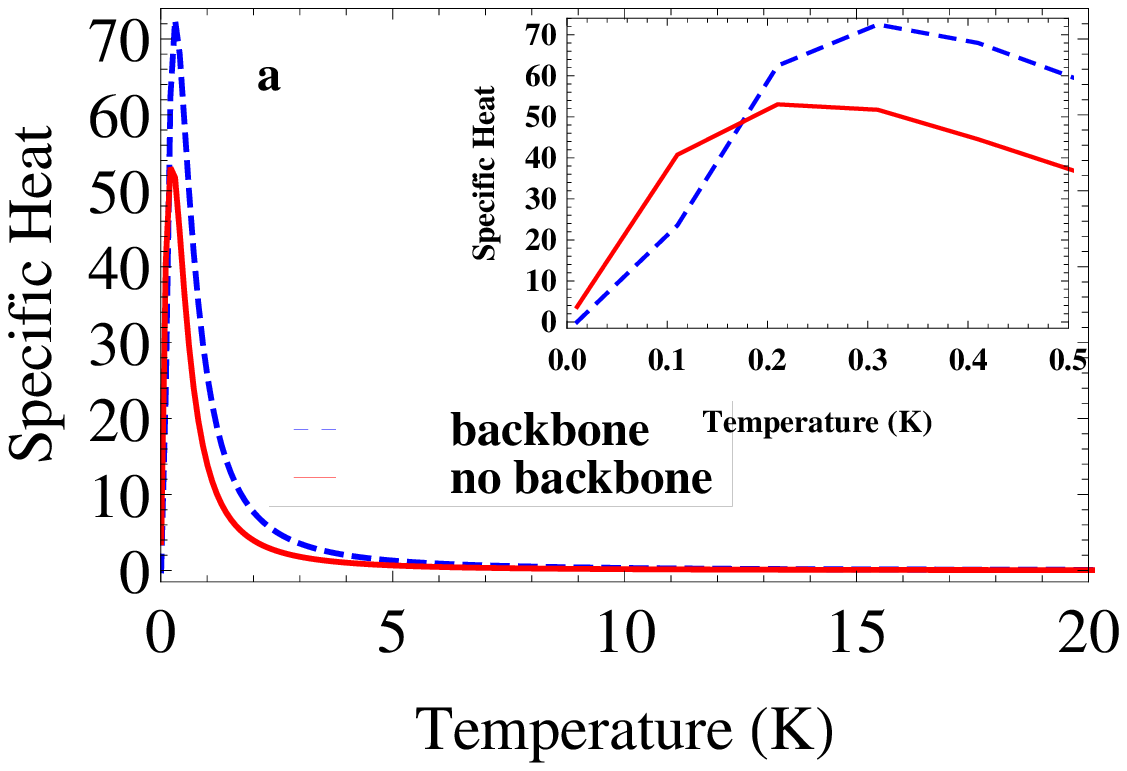}\\
     \includegraphics[width=60mm]{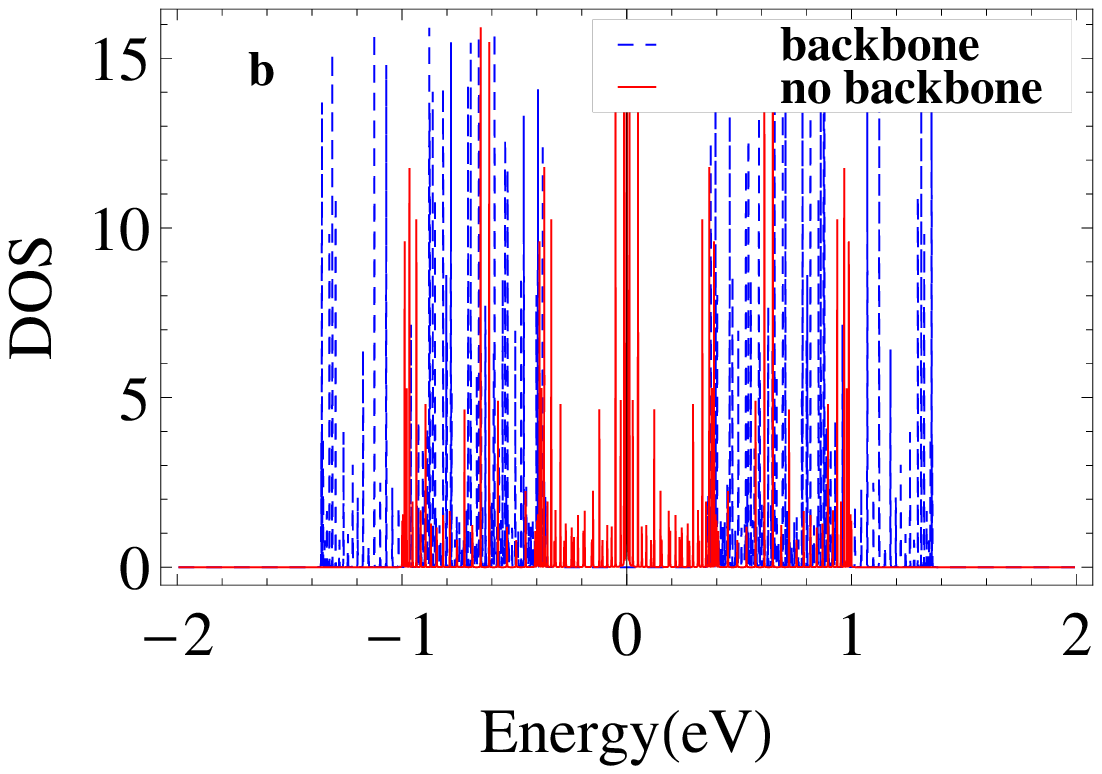}
  \end{tabular}
\caption{(Color online) a. Electronic specific heat $(C_v)$ vs temperature ($T$) 
plot with and without backbone (no sequence, all site-energies are set to zero). 
$C_v$ increases as we introduce backbone into the system except at low 
temperature (see inset). b. Density of states (DOS) vs. energy (E) with and 
without backbone. It is clearly visible that due to introduction of backbone 
a gap opens in the spectrum.}

\label{fig2}
\end{figure}

 In Fig.~\ref{fig2} we show the behavior of specific heat ($C_v$) with 
temperature ($T$) for a simple ladder model and also for the DBLM which 
incorporates the backbone structure of the DNA. For the sake of comparisons 
of both the models we set the site-energies of the nucleotides and the backbone 
sites to zero. Here we have no sequencing of DNA and also we have ignored disorder 
( {\it i.e.}, w=0) due to environmental fluctuations.  It is clear from the 
figure that due to the presence of backbone,  $C_v$ gets increased at almost all 
the temperatures, excepting a small low-temperature region. The corresponding DOS 
profiles are also shown alongside, which reveals that a gap opens up in presence 
of backbones. 

\begin{figure}[ht]
\centering

  \begin{tabular}{cc}
   \includegraphics[width=35mm, height=28mm]{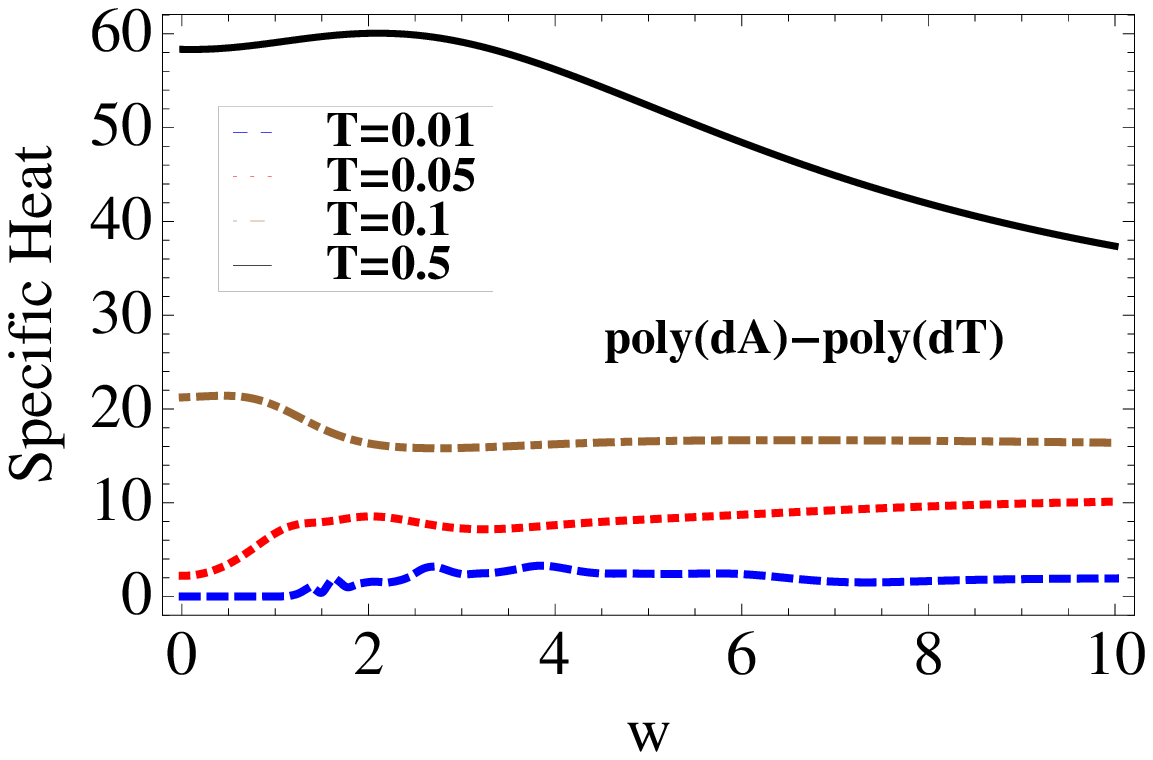}&
    \includegraphics[width=35mm, height=28mm]{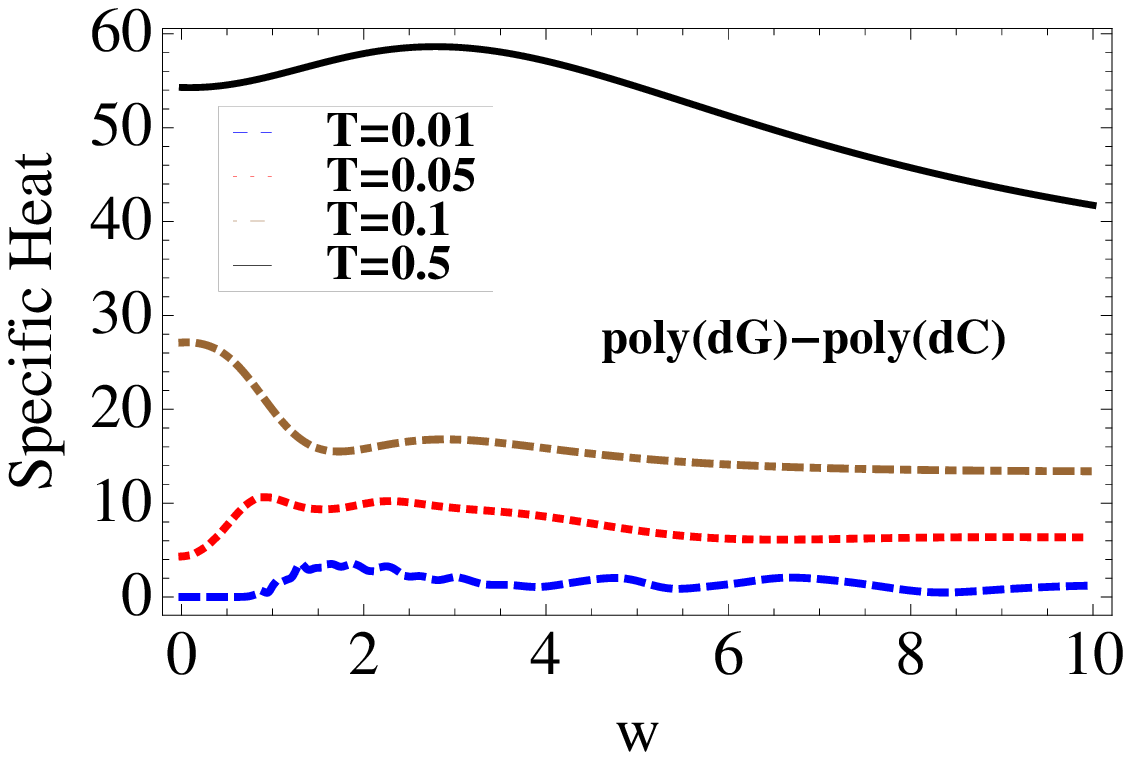}\\

     \includegraphics[width=35mm, height=28mm]{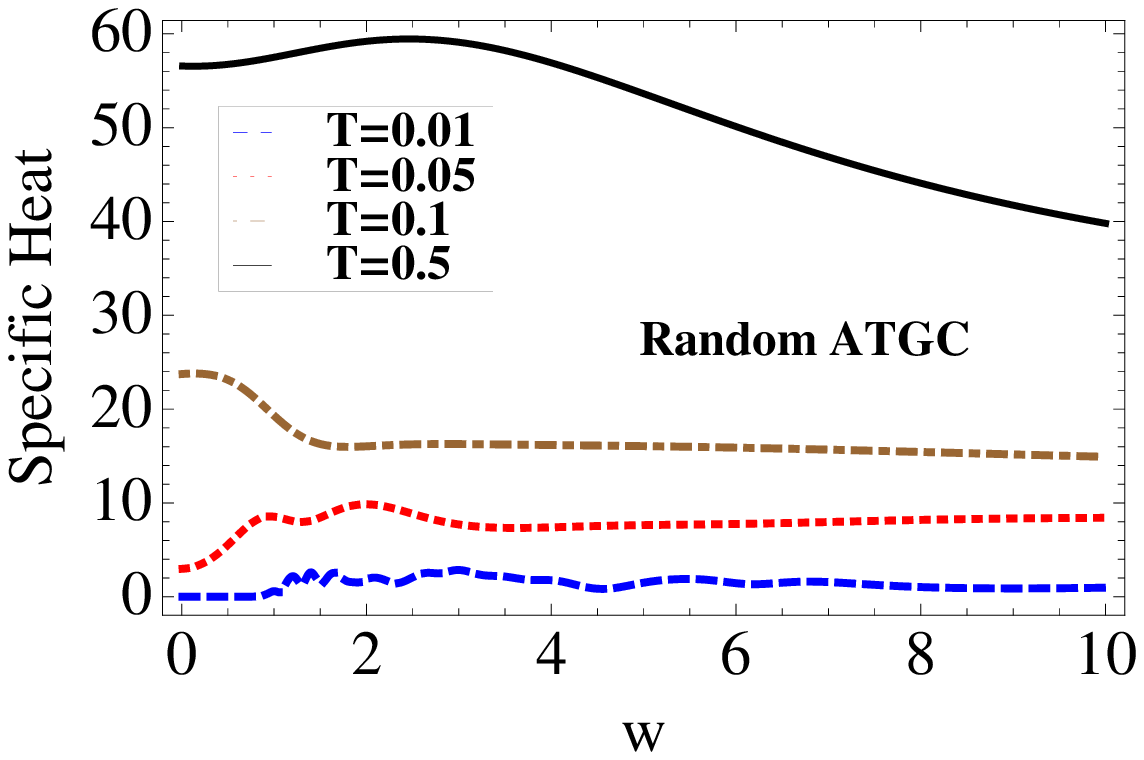}&
      \includegraphics[width=35mm, height=28mm]{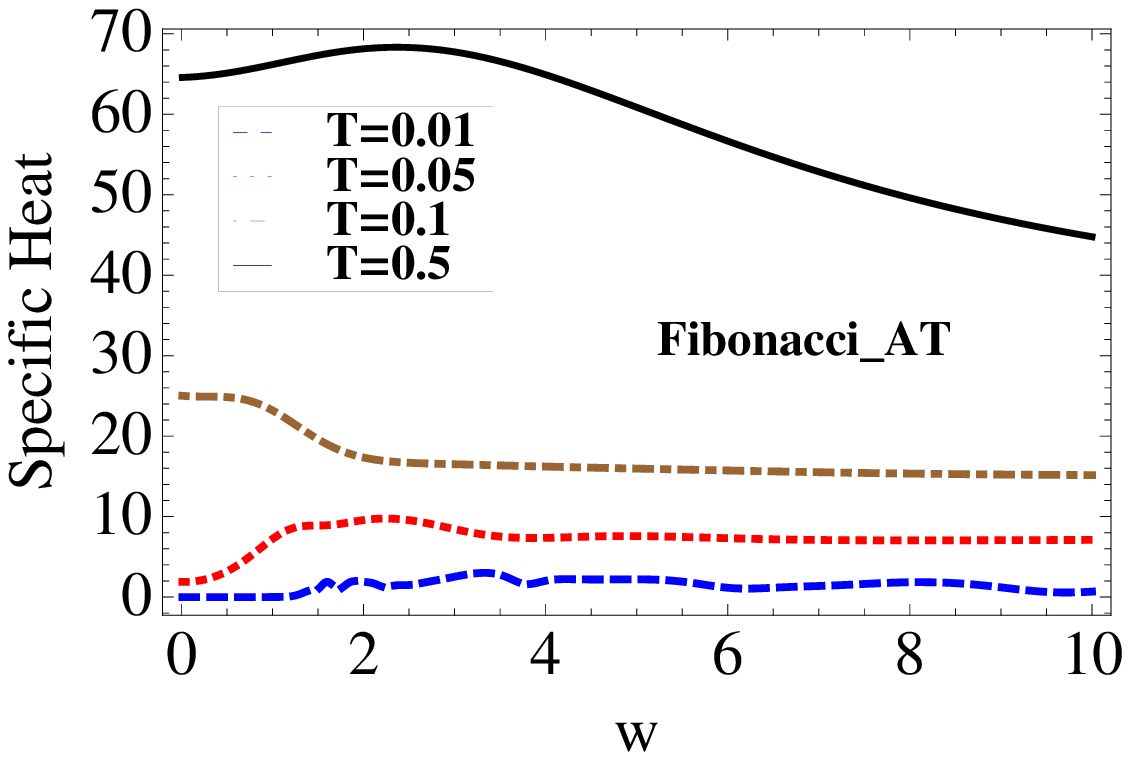}
  \end{tabular}

\caption{(Color online) Electronic specific heat $(C_v)$ as a function of 
backbone disorder strength (w) for four different DNA sequences at low 
temperature ($T<$2K). $C_v$ decreases with disorder (w) except initial 
rise at low w for all the cases.} 

\label{fig3}
\end{figure}

First let us explain the basic nature of the $C_v$ vs temperature ($T$) 
curve. At low temperature only the states within the range $E_F\pm kT$ are 
accessible to the electrons, with $E_F$ being the Fermi energy. As the 
average energy of the system is given by $<E>$ = $\int E \rho(E) f(E) dE $, 
at low temperature we can make the following approximations: dE~$\approx$ kT, 
$\rho(E)\approx\rho$~= a constant, f(E)~$\approx$1, and the average 
energy becomes $<E>~\approx\rho k^2T^2$. Then specific heat becomes $C_v$~= $\rho k^2T$, 
being proportional to the temperature, and $C_v$ will increases with temperature 
at low temperature. Let us now see the high temperature behaviour of DNA. 
The DNA system is a finite one, its energy spectra forms a band of finite 
width and at high temperature all these states are accessible to the electrons. 
Now as we increase the temperature, in the very high temperature limit 
almost all the states are equally populated and the average energy $<E>$ 
becomes almost independent of temperature. So as we increase temperature 
from low temperature regime, specific heat initially increases with temperature 
and then it decreases with temperature and finally goes to zero in the very 
high temperature regime. 

\begin{figure*}
\centering

 \begin{tabular}{cccc}
 
  \includegraphics[width=35mm, height=27mm]{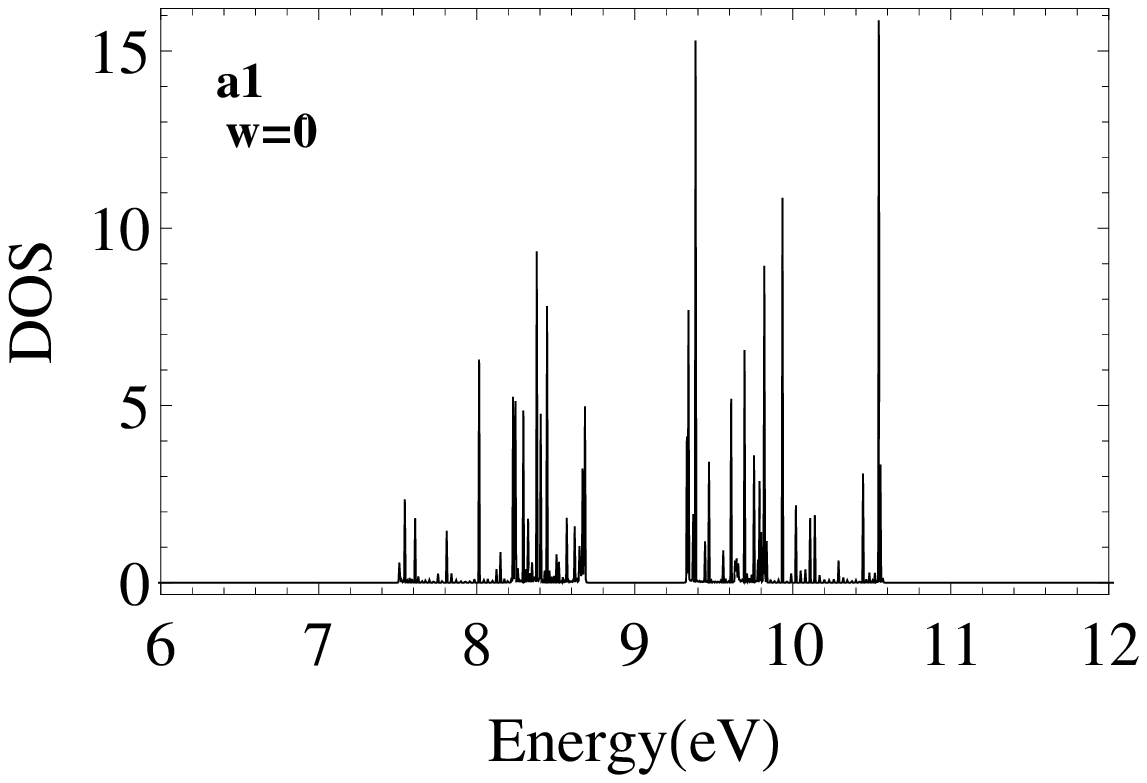}&
   \includegraphics[width=35mm, height=27mm]{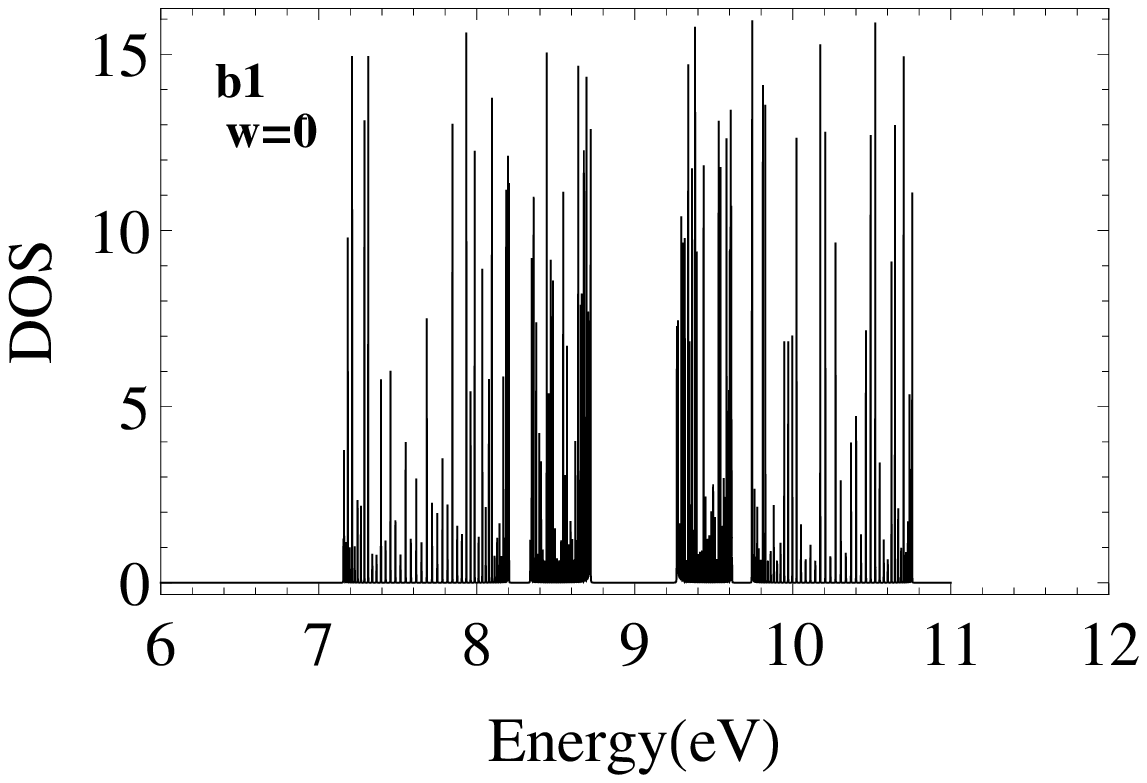}&
    \includegraphics[width=35mm, height=27mm]{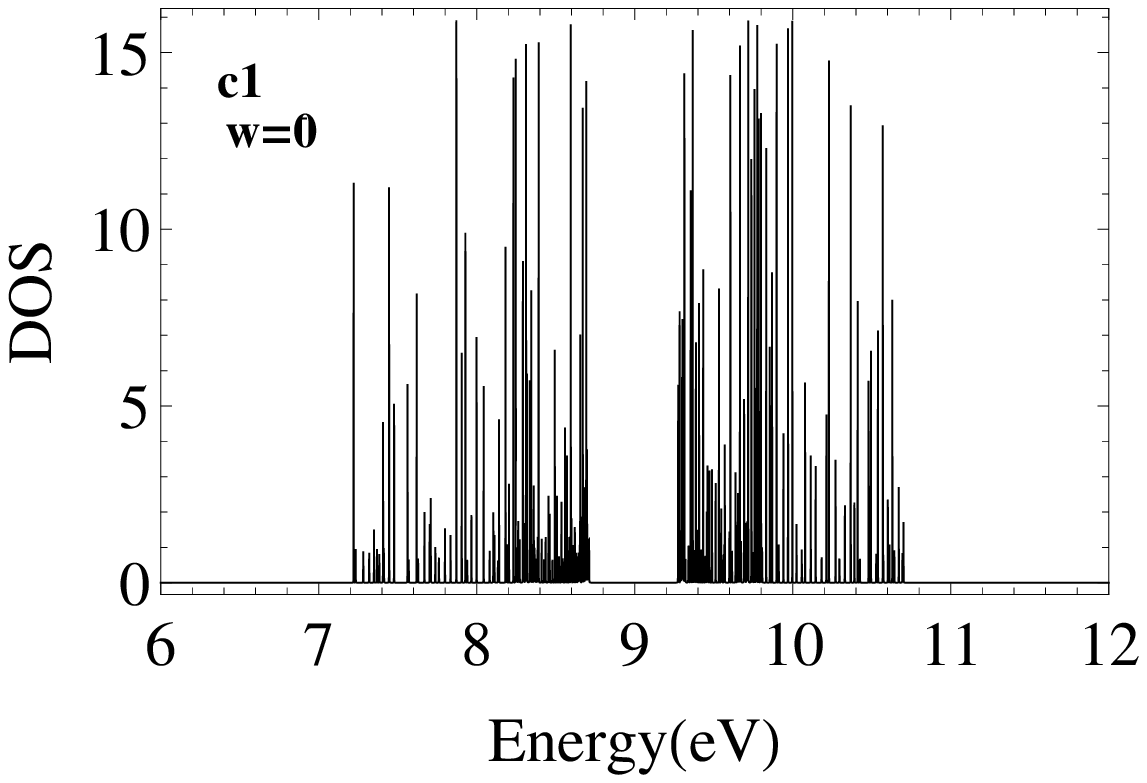}&
     \includegraphics[width=35mm, height=27mm]{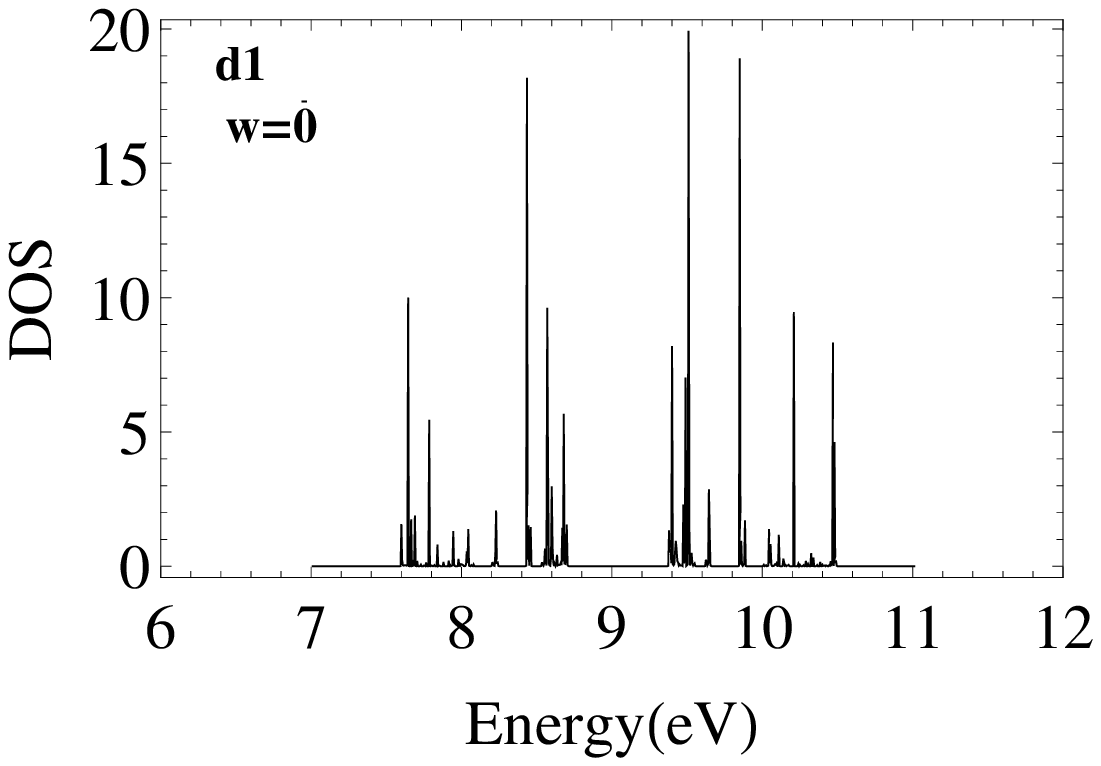}\\
  
   \includegraphics[width=35mm, height=27mm]{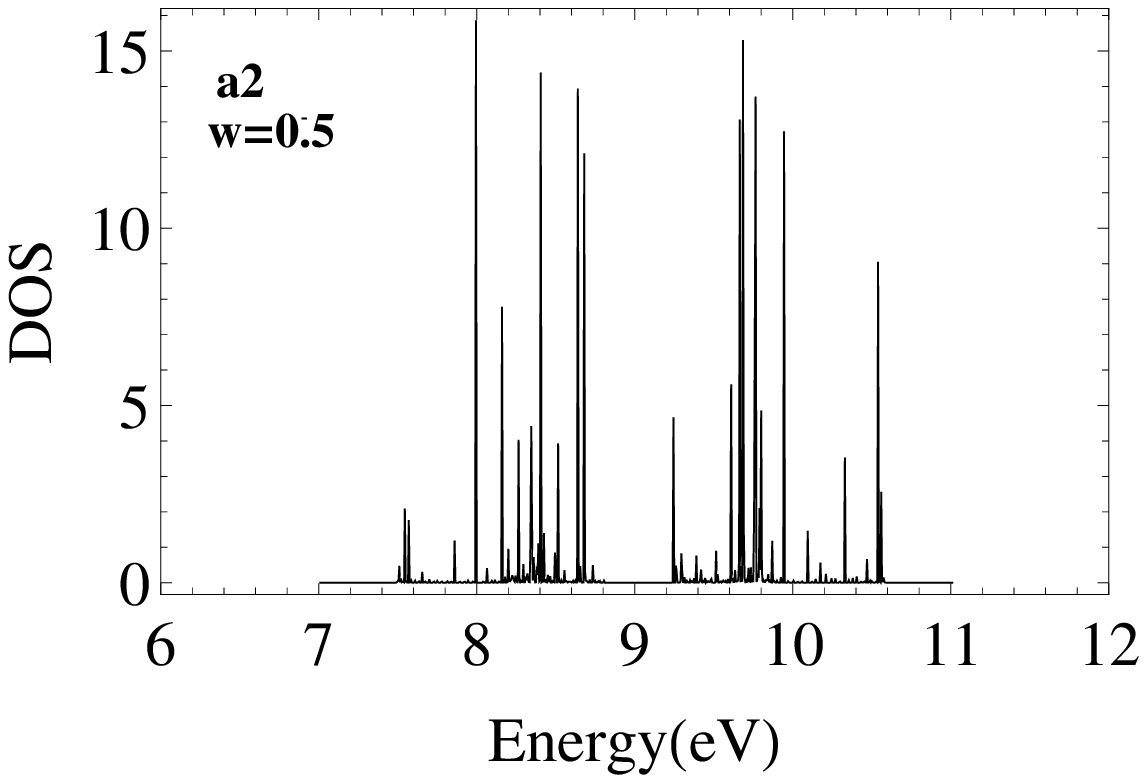}&
    \includegraphics[width=35mm, height=27mm]{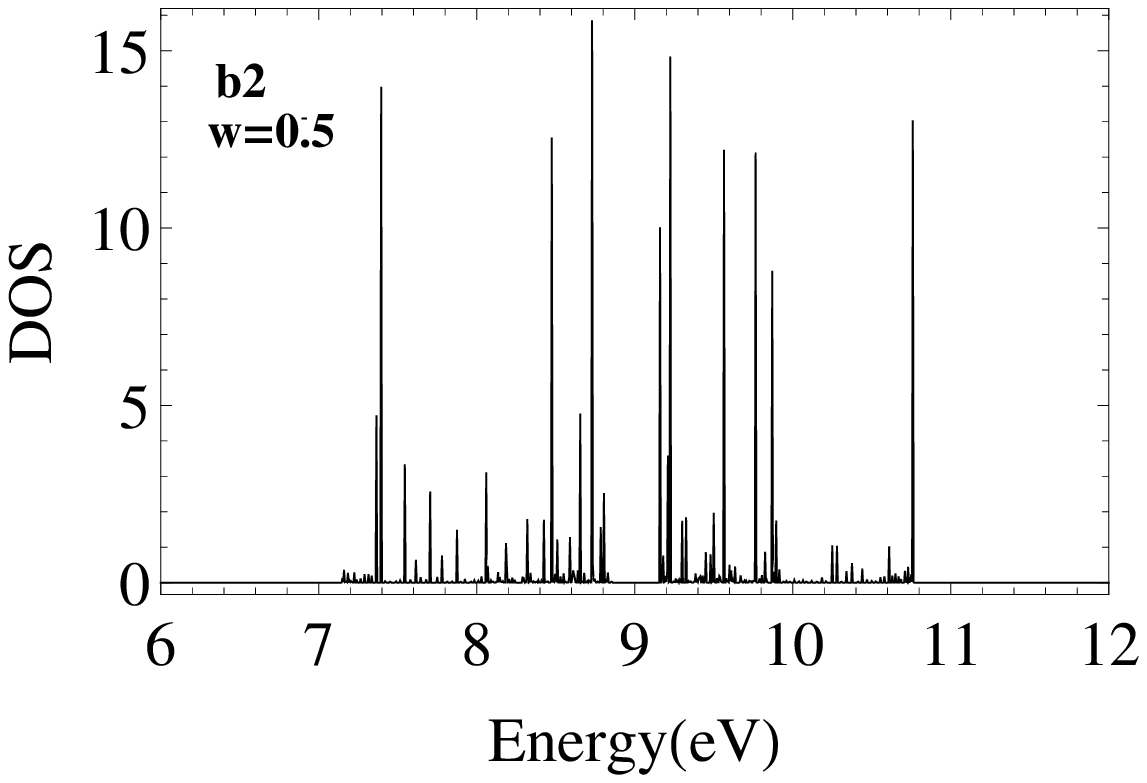}&
     \includegraphics[width=35mm, height=27mm]{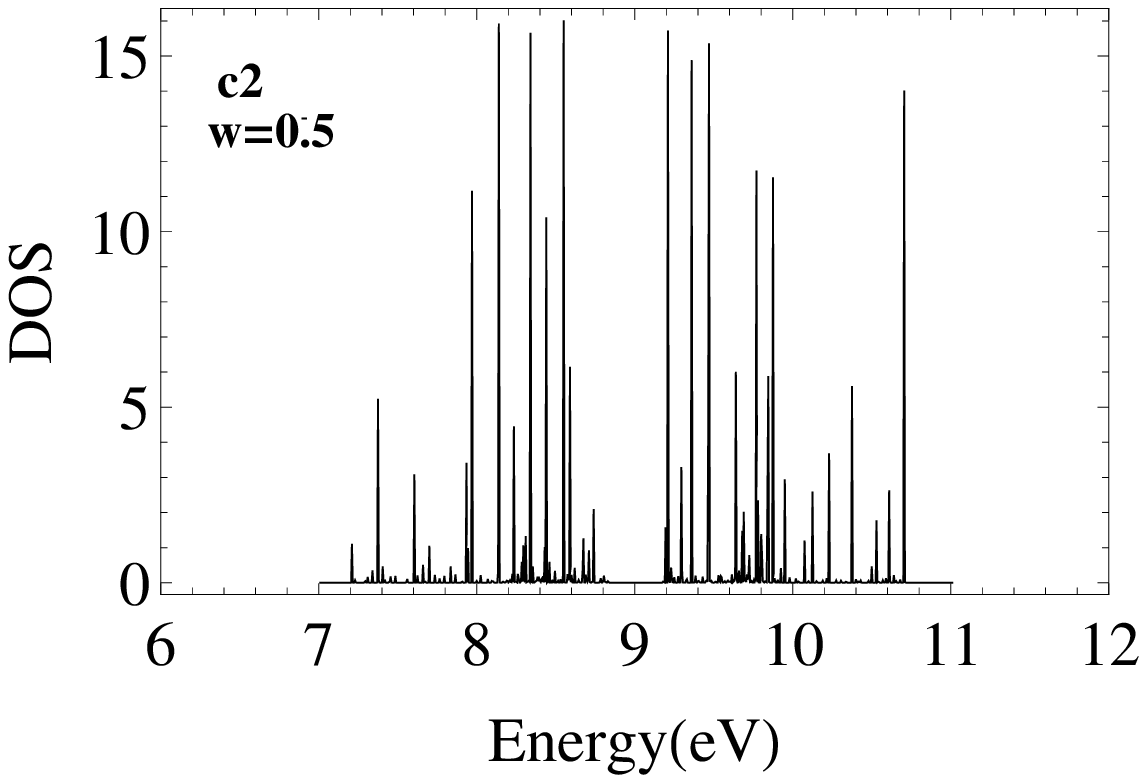}&
      \includegraphics[width=35mm, height=27mm]{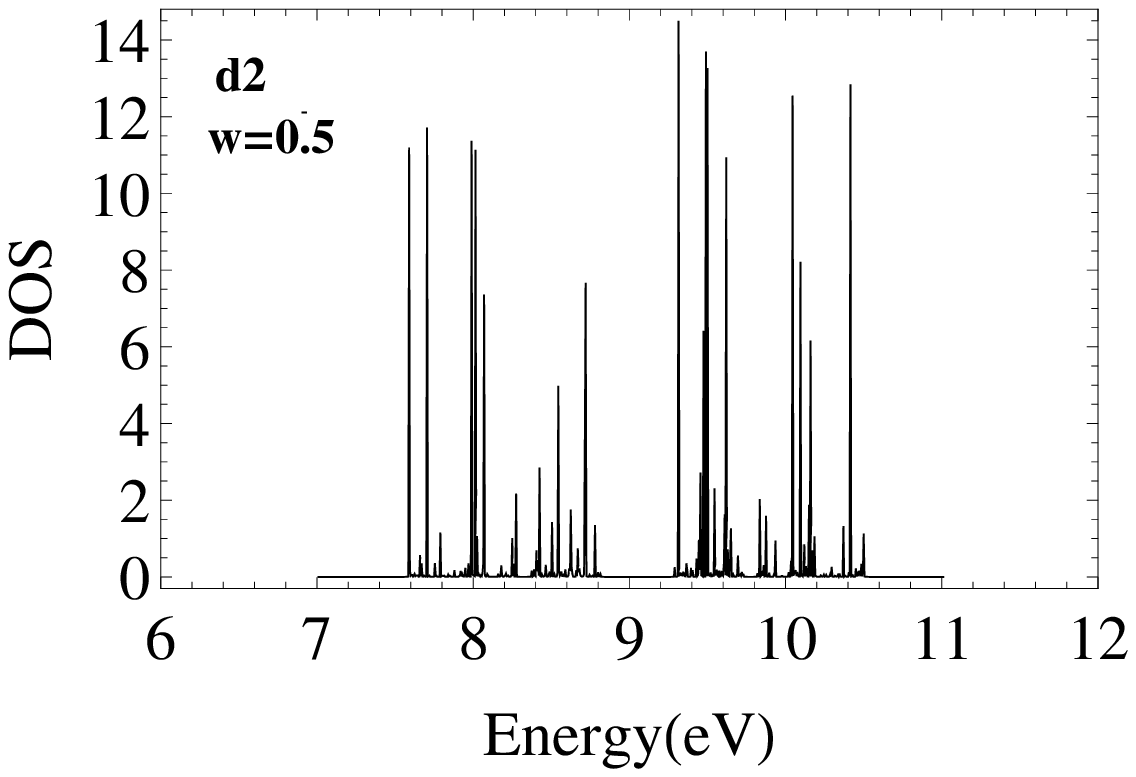}\\
   
    \includegraphics[width=35mm, height=27mm]{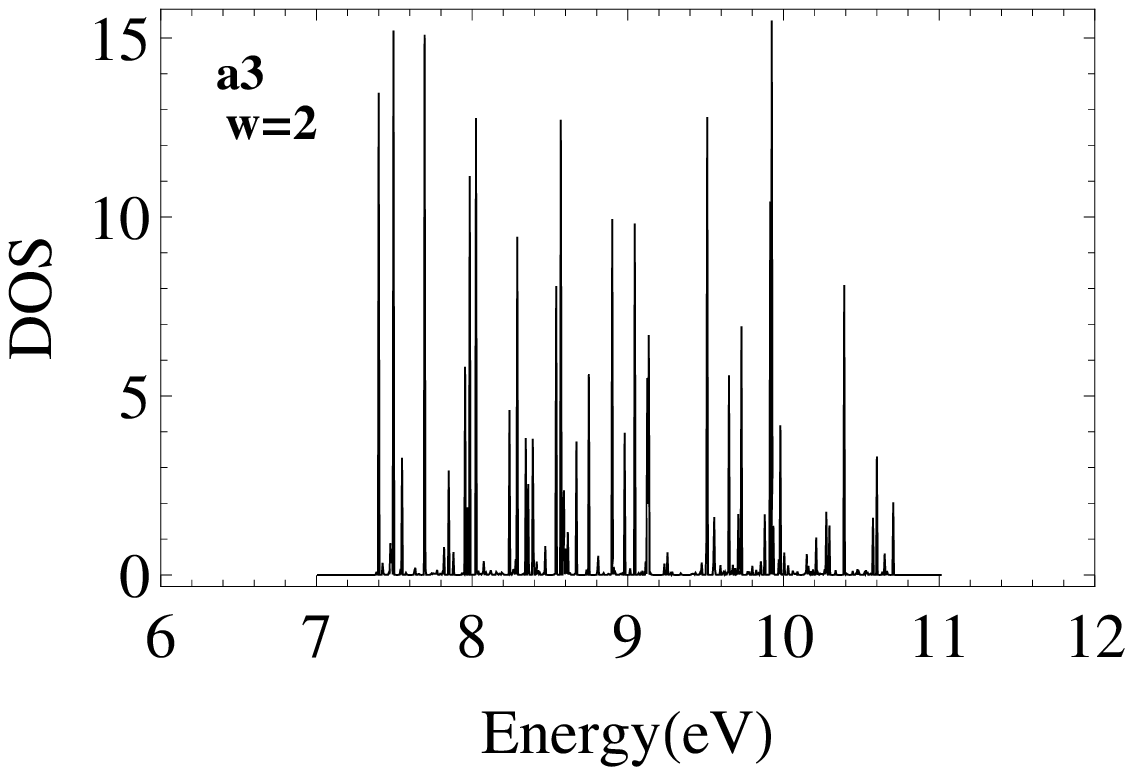}&
     \includegraphics[width=35mm, height=27mm]{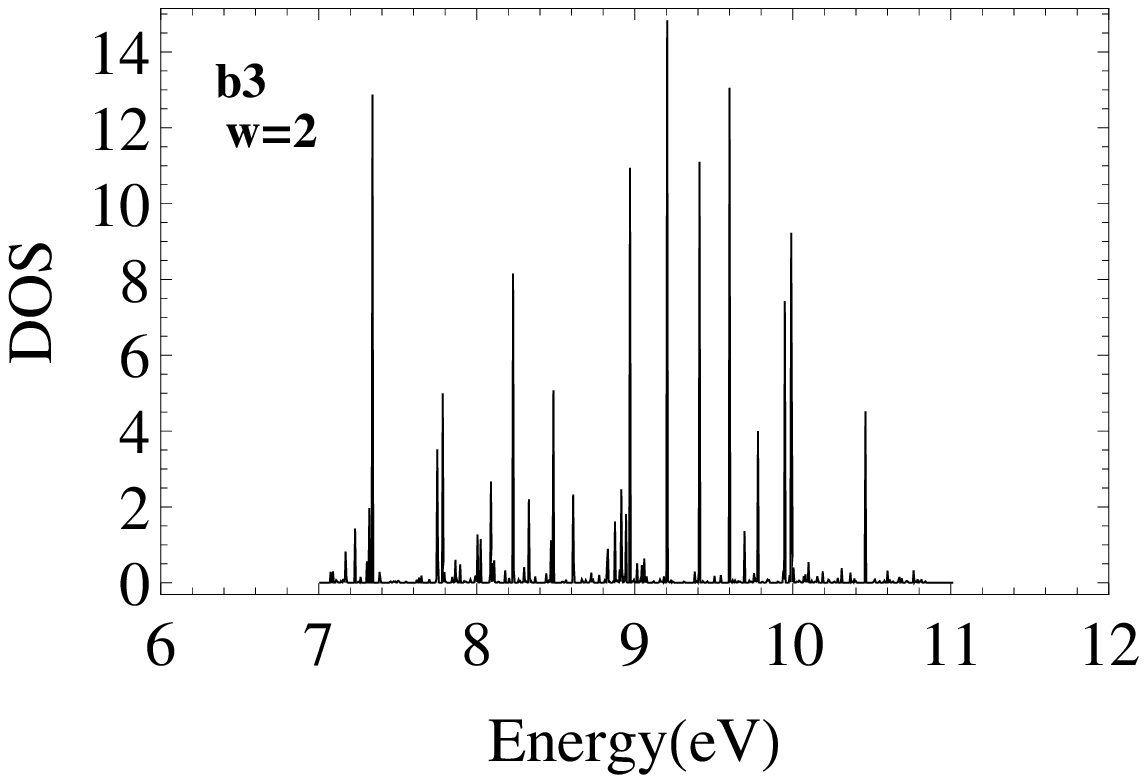}&
      \includegraphics[width=35mm, height=27mm]{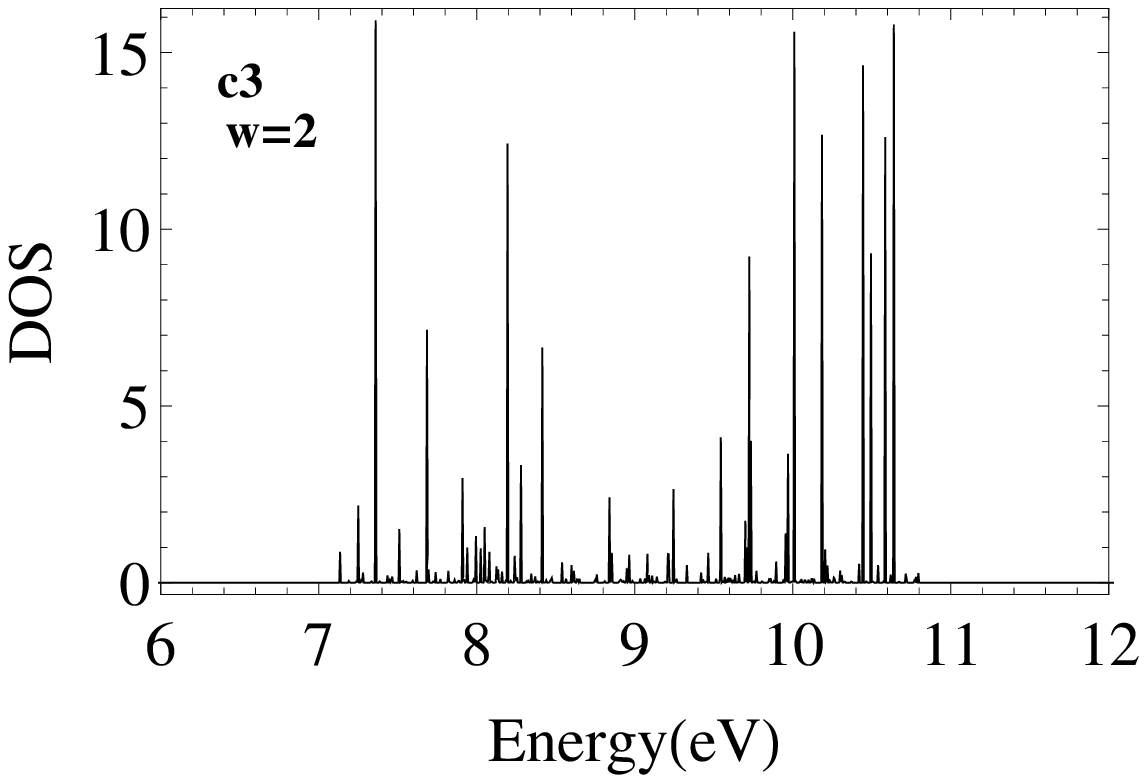}&
       \includegraphics[width=35mm, height=27mm]{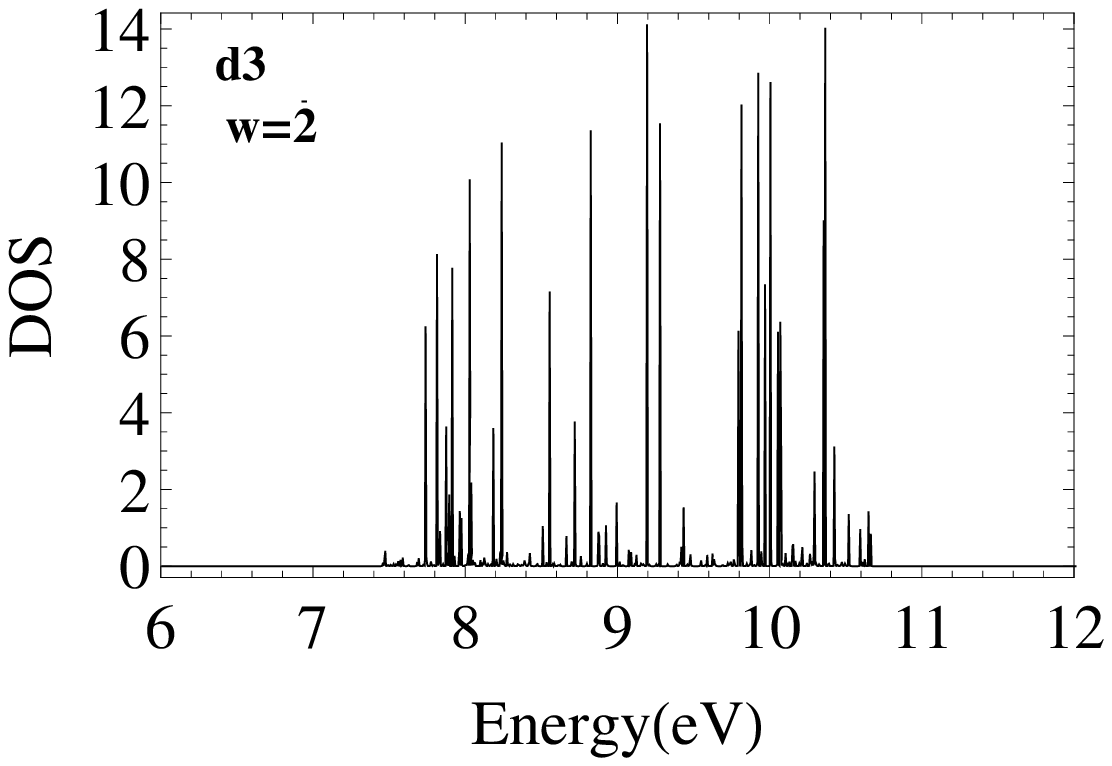}\\
    
     \includegraphics[width=35mm, height=27mm]{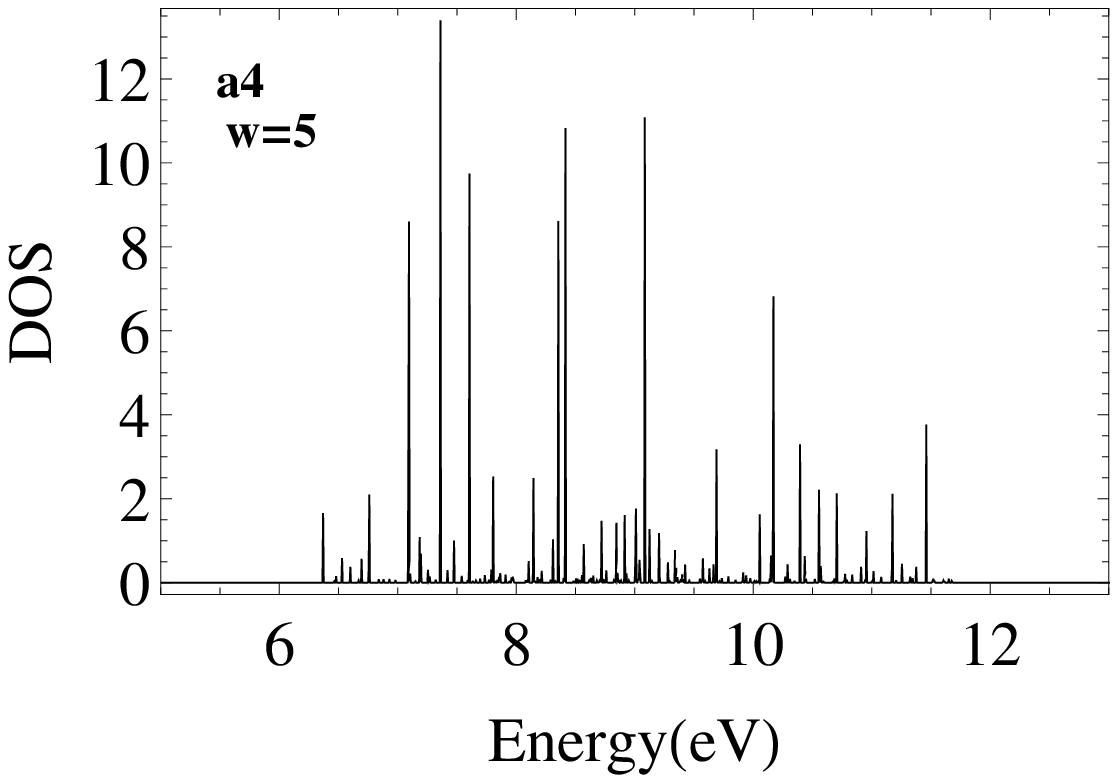}&
      \includegraphics[width=35mm, height=27mm]{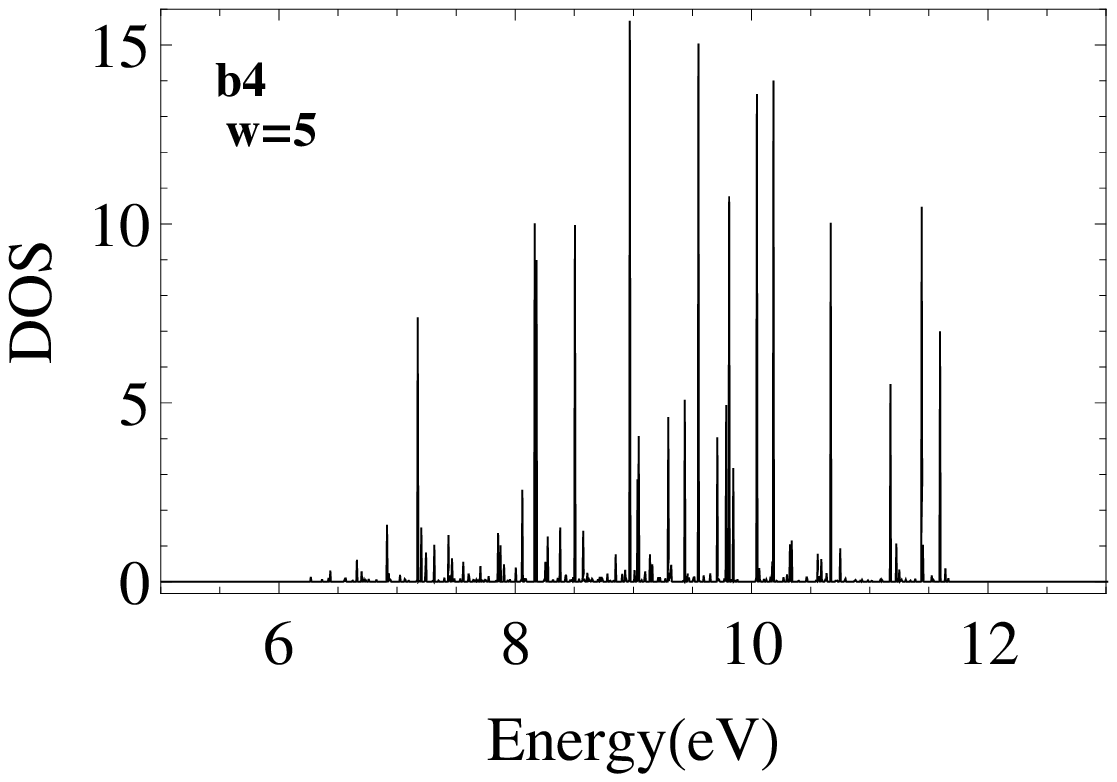}&
       \includegraphics[width=35mm, height=27mm]{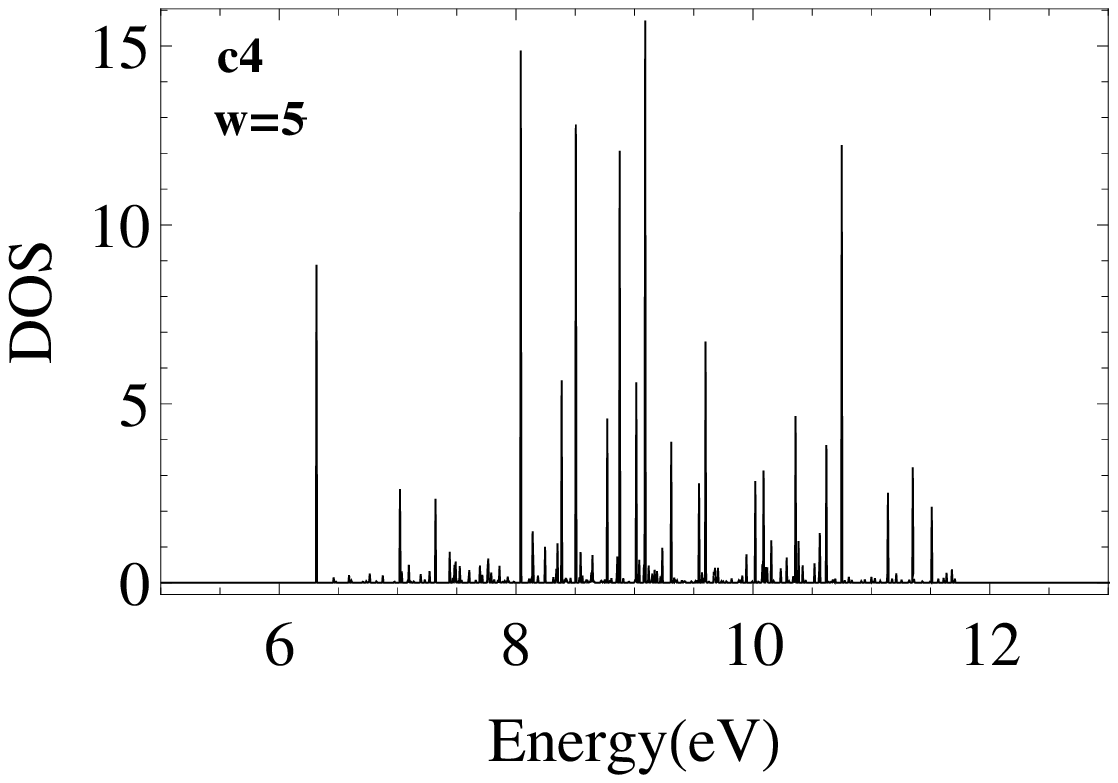}&
        \includegraphics[width=35mm, height=27mm]{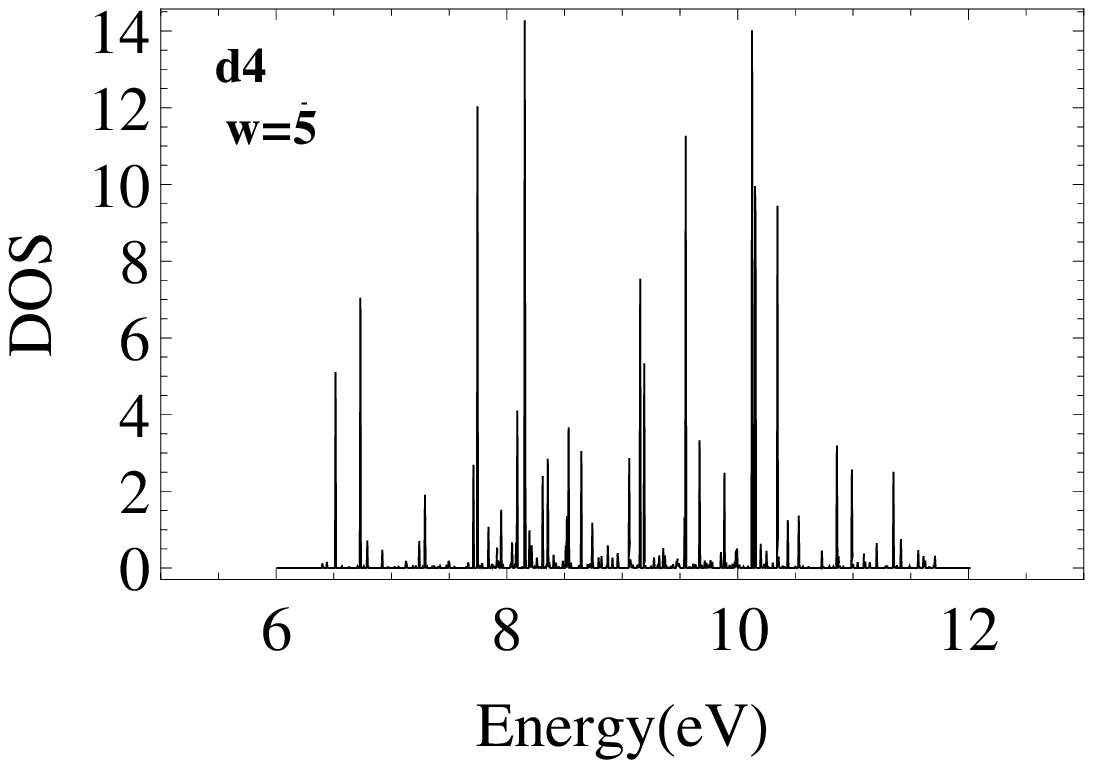}\\
     
      \includegraphics[width=35mm, height=27mm]{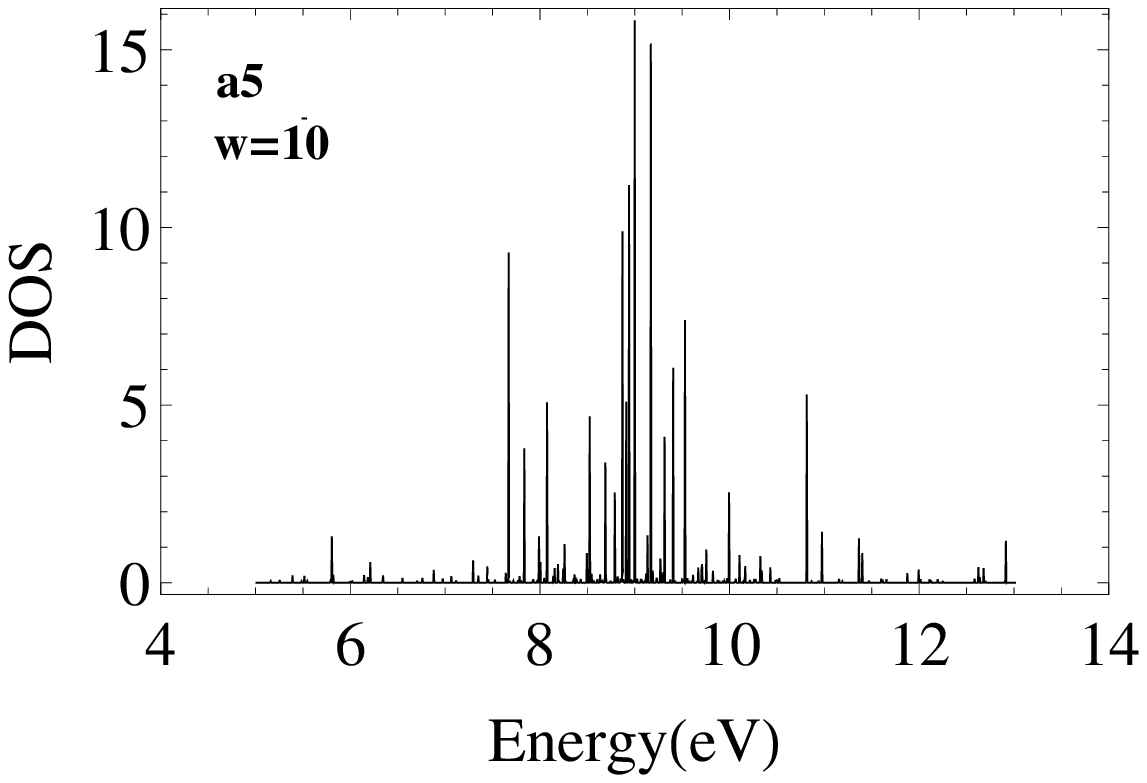}&
       \includegraphics[width=35mm, height=27mm]{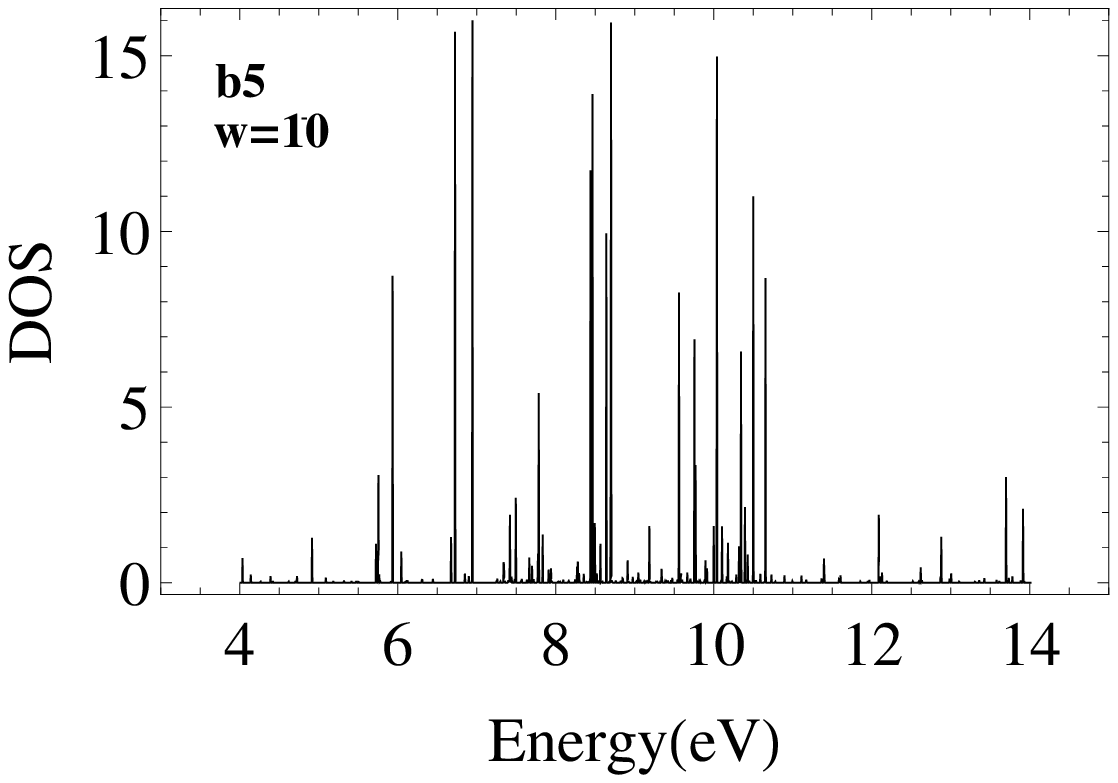}&
        \includegraphics[width=35mm, height=27mm]{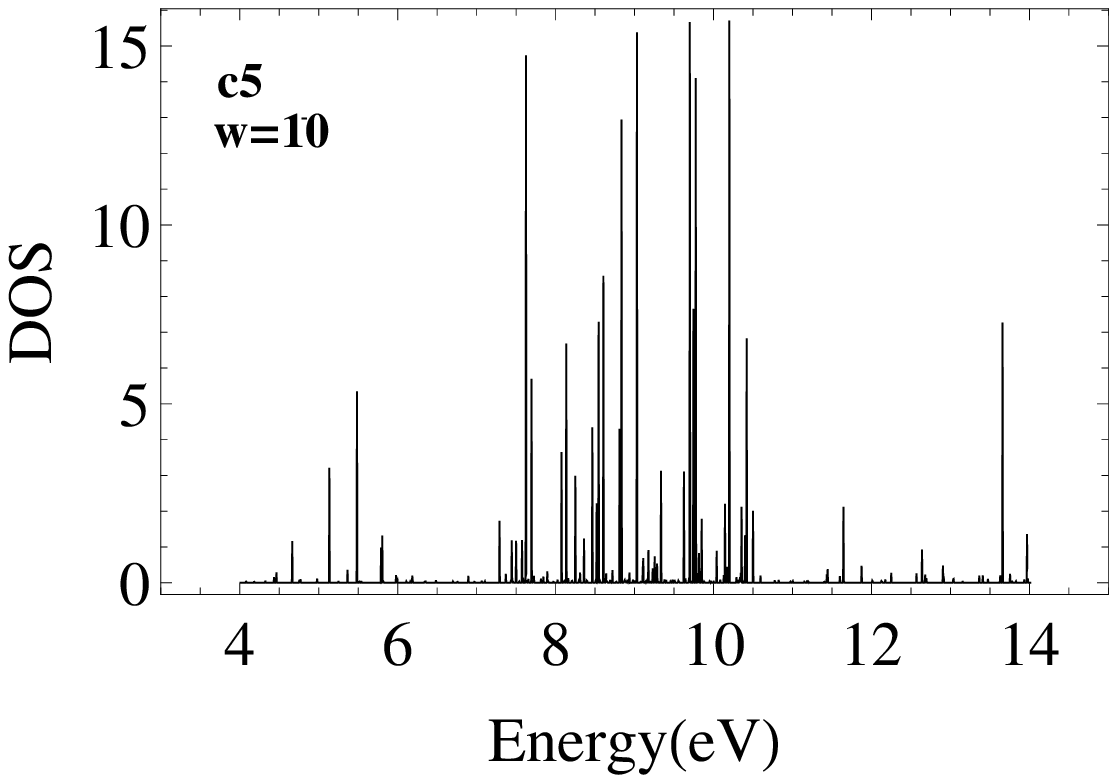}&
         \includegraphics[width=35mm, height=27mm]{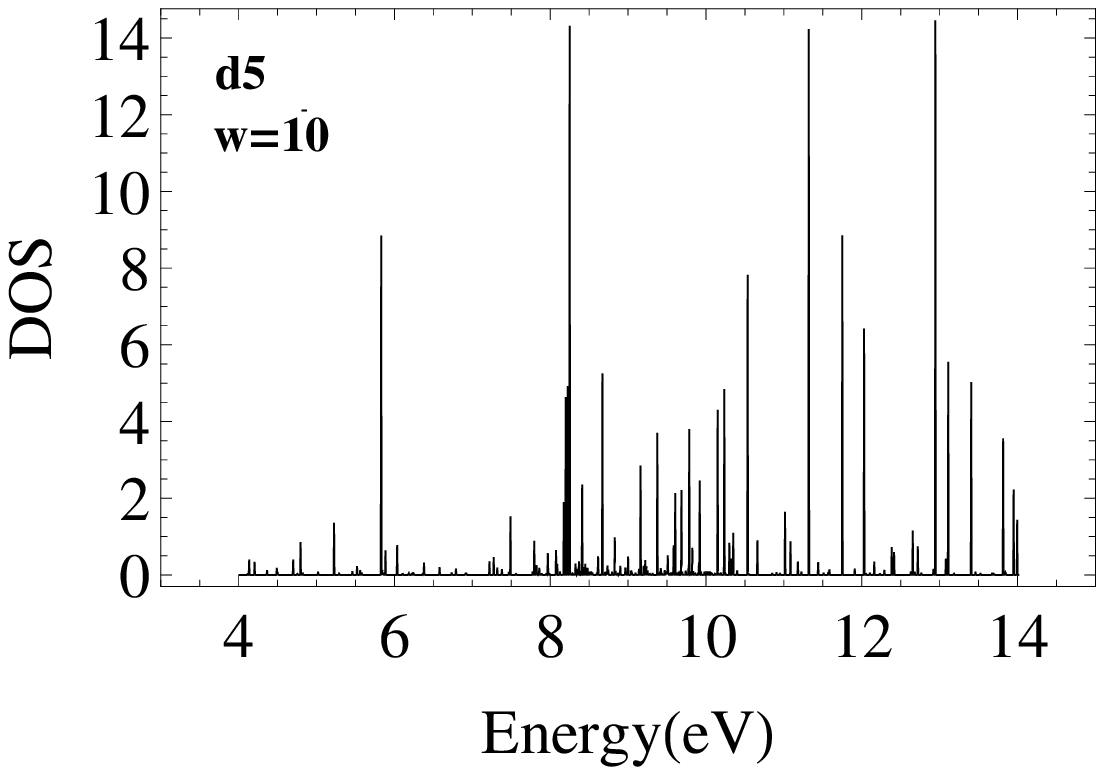}\\
      
       \includegraphics[width=35mm, height=27mm]{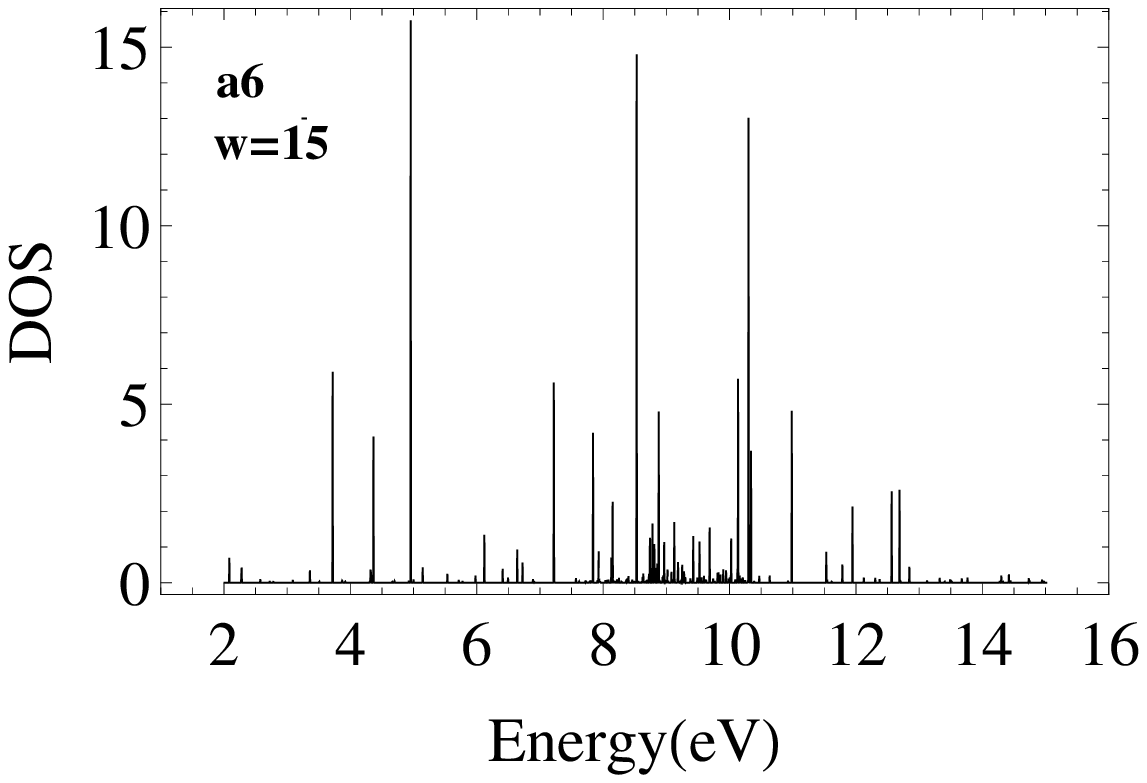}&
        \includegraphics[width=35mm, height=27mm]{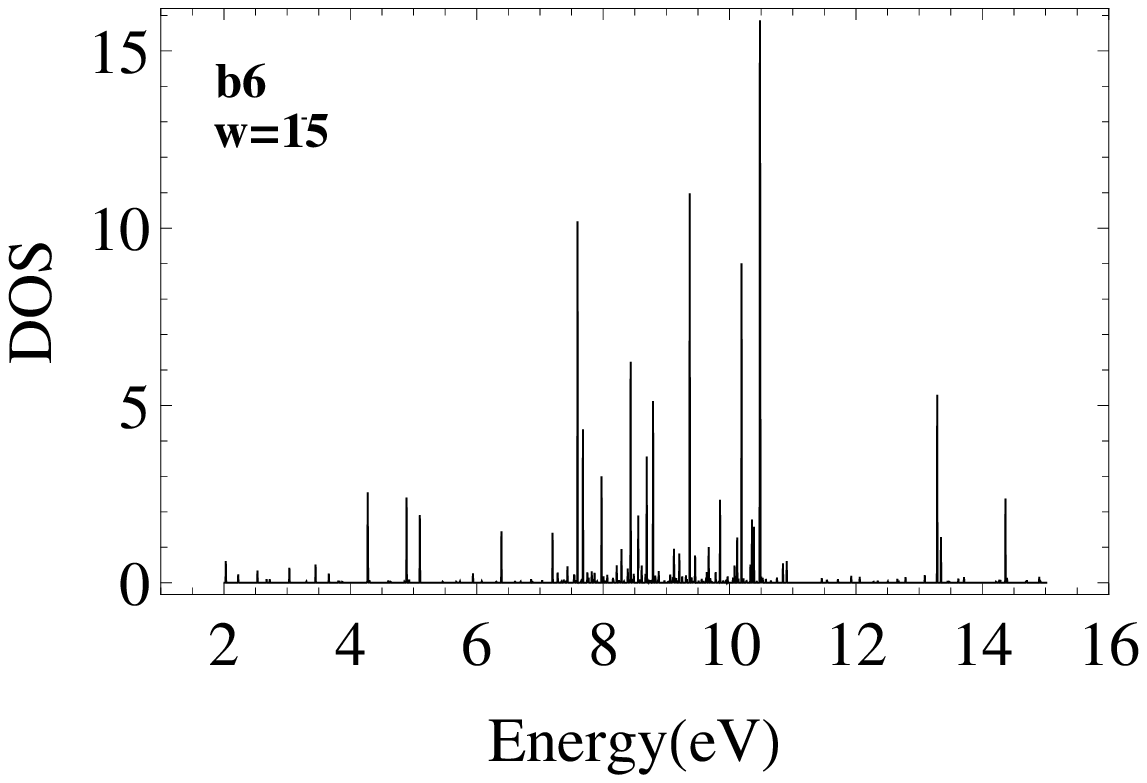}&
         \includegraphics[width=35mm, height=27mm]{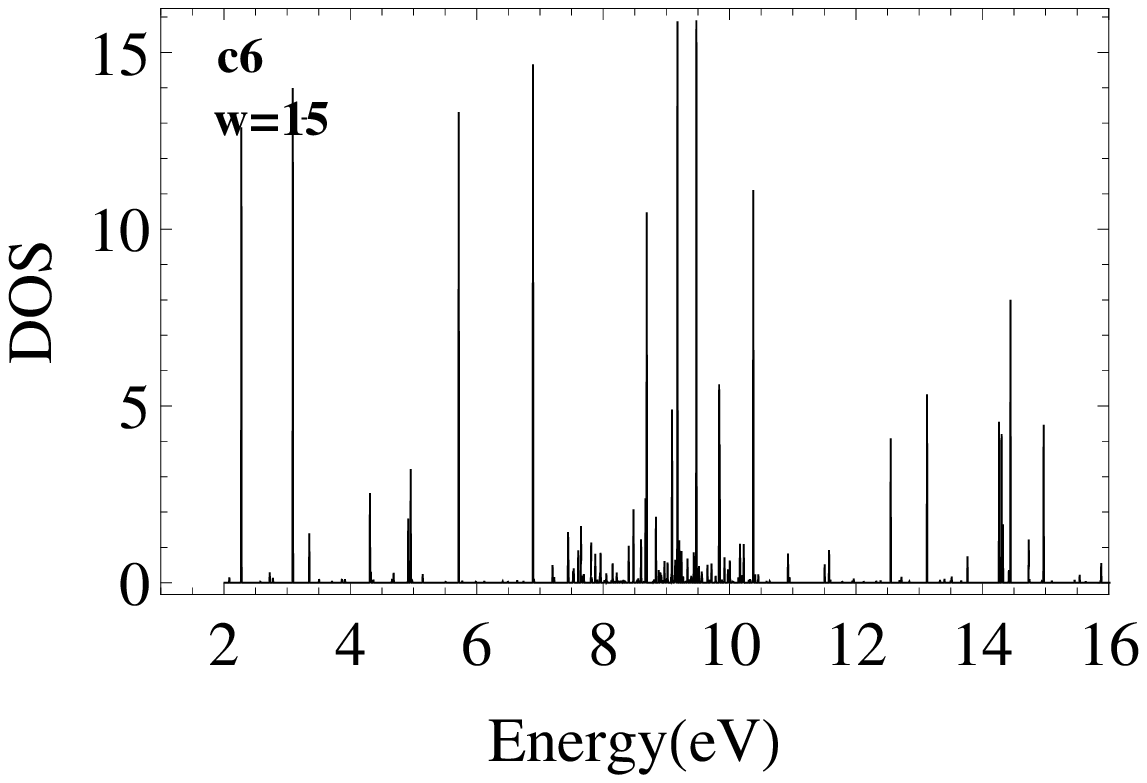}&
          \includegraphics[width=35mm, height=27mm]{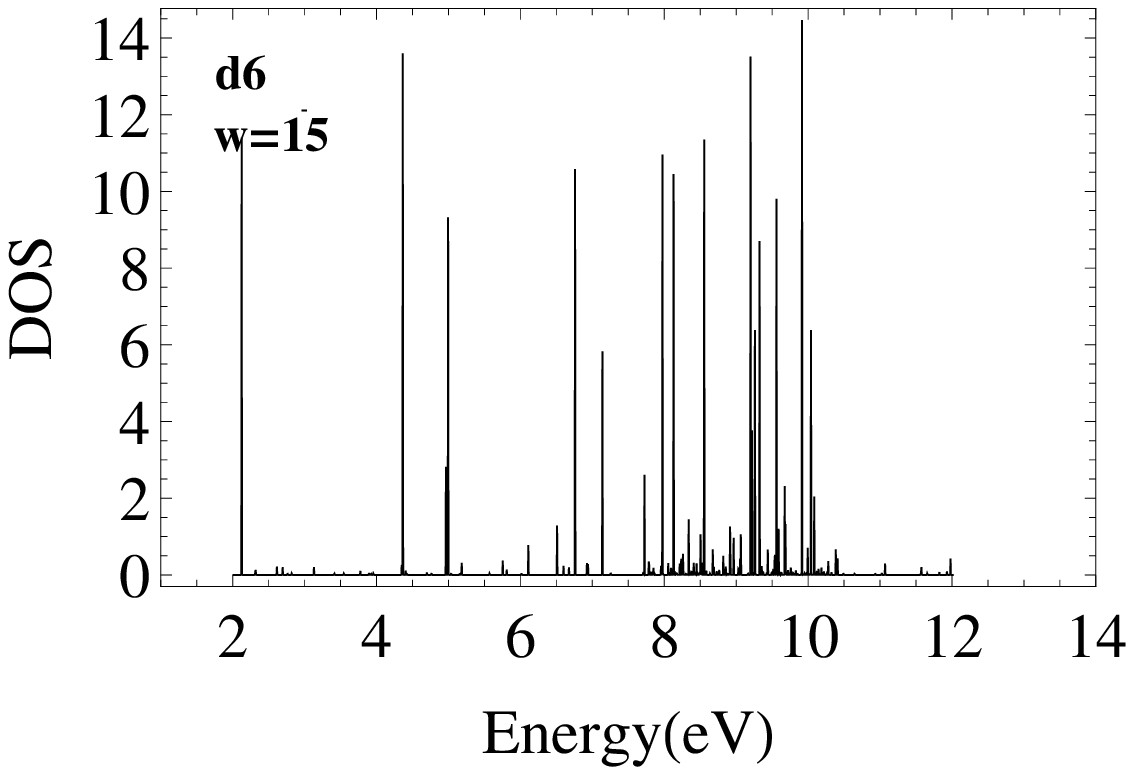}\\     
       
        \includegraphics[width=35mm, height=27mm]{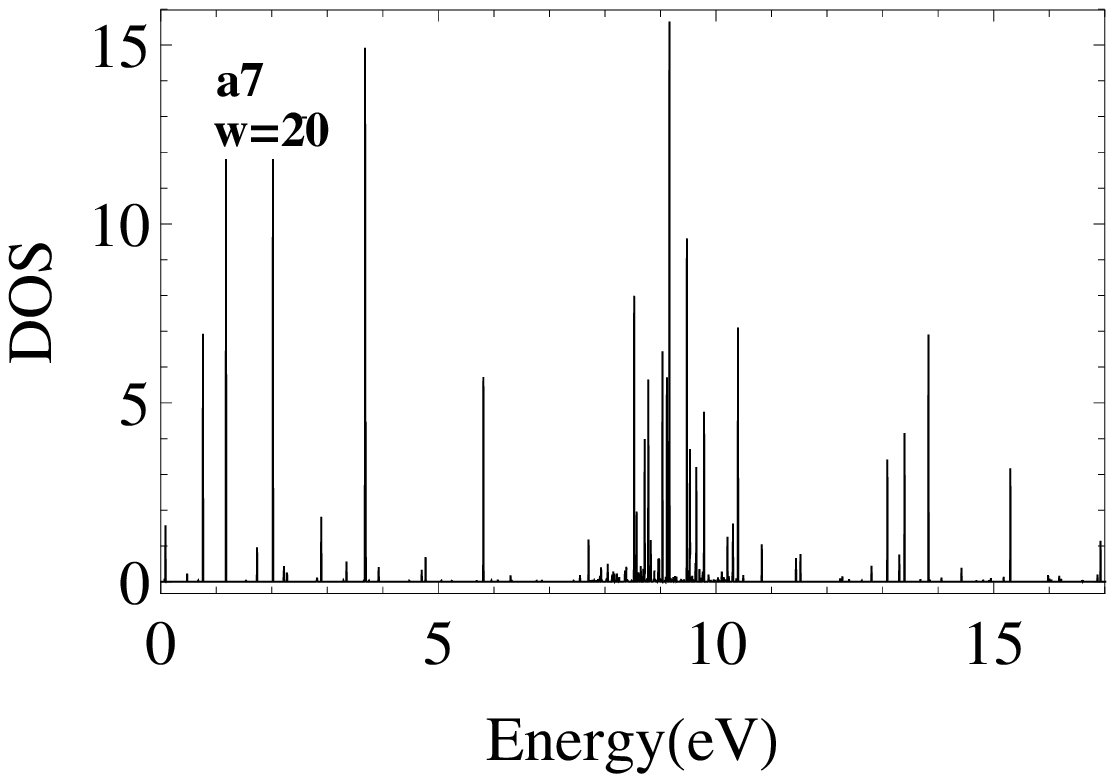}&
         \includegraphics[width=35mm, height=27mm]{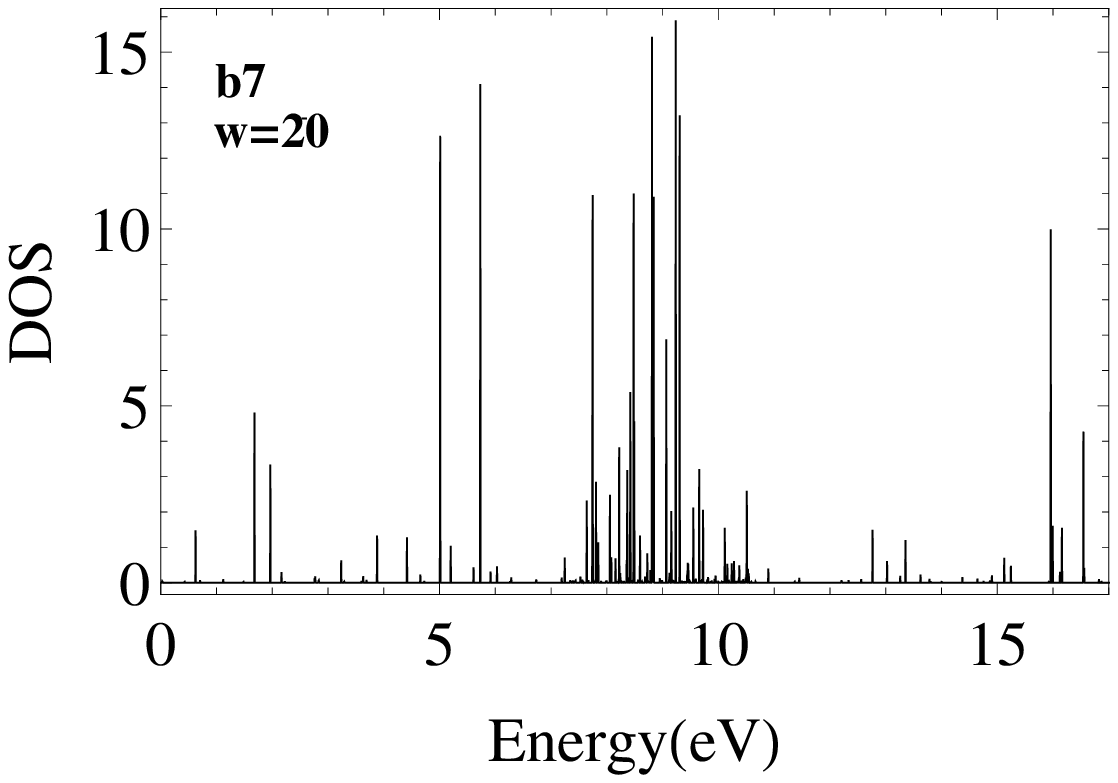}&
          \includegraphics[width=35mm, height=27mm]{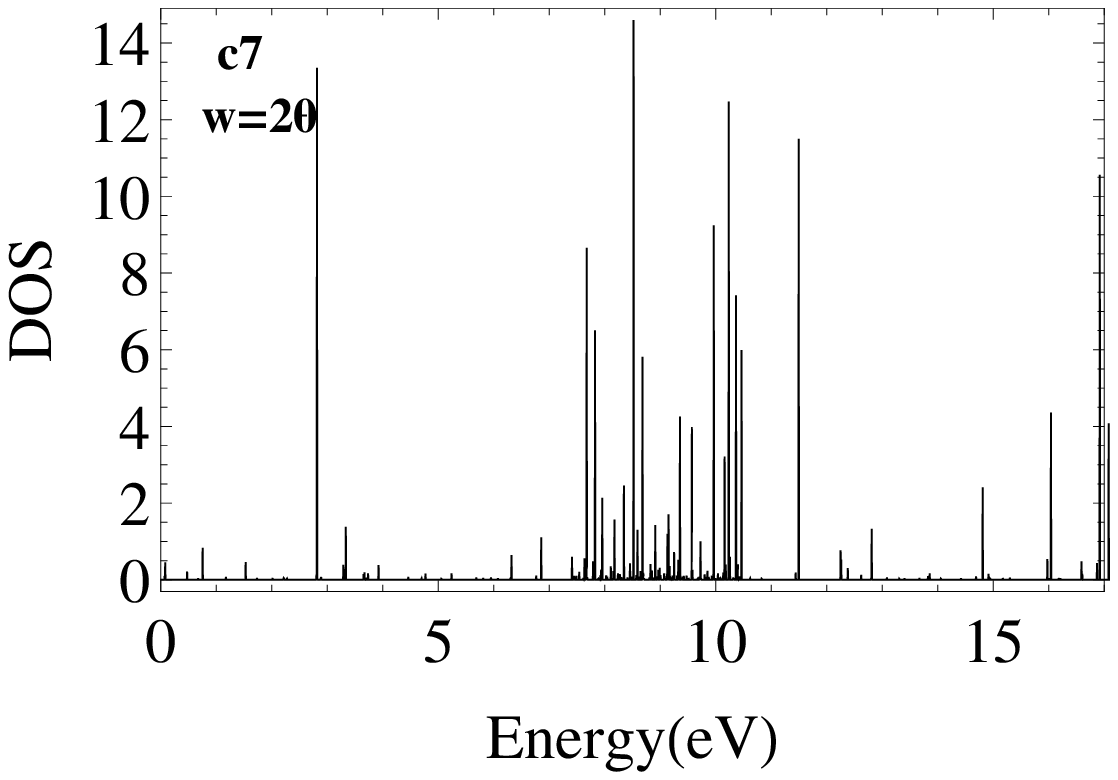}&
           \includegraphics[width=35mm, height=27mm]{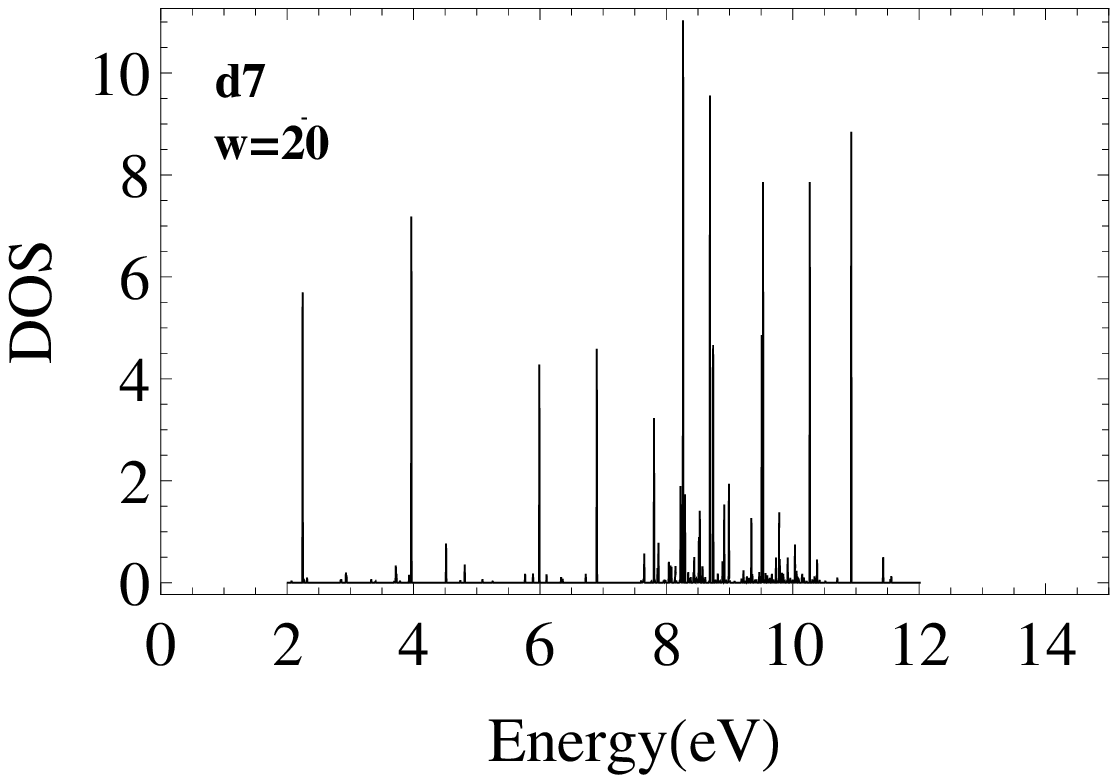}\\
              
  \end{tabular}

\caption{(Color online). Density of states (DOS) profiles for four DNA 
sequences at different disorder strengths (w). a1-a7 are poly(dA)-poly(dT),  
b1-b7 are poly(dG)-poly(dC), c1-c7 are random ATGC and d1-d7 are Fibonacci 
sequences for disorder strength w= 0, 0.5, 2, 5, 10, 15, 20 respectively 
along vertical lines.}

\label{fig4}
\end{figure*}

Following similar arguments we can now explain the behavior of the 
result of Fig.~\ref{fig2}. A gap opens up in the middle of the 
spectrum due to presence of the backbones, and we have a gap at Fermi 
level in the half filled case. With increase of temperature at the low 
temperature range initially no states are accessible to the electrons. 
Now at sufficiently high temperature, when thermal energy becomes 
comparable to the energy gap of the system, the excited states of 
the DBLM gradually become accessible to the electrons despite of 
the energy gap. Thus $<E>$ of the system increases with T at a slower 
rate than the simple ladder model without backbone since it has no gap 
in the spectrum. So, initially at the low temperature regime $C_v$ gets 
lowered due to introduction of the backbones (see inset of Fig. 2). 
As the excited states of DBLM has energy higher than that 
of the simple ladder model, so $<E>$ will increase at a much 
higher rate at high temperature and accordingly $C_v$ will be 
higher due to presence of backbones in the high temperature 
regime. 

 In Fig.~\ref{fig3} we plot the variation of specific heat with 
backbone disorder in the low temperature range T$<$2K. It is observed 
from the figure that $C_v$ increases first at low disorder and then it 
decreases monotonically as the disorder strength increases. The reason is 
clear from the DOS profiles provided in Fig.~\ref{fig4}. 
At zero disorder (w=0) there is gap in the system for all the sequences, 
for small w new states started to appear near around the Fermi energy 
$(E_F)$ and the gap started to diminish. These new states can be accessed 
at low temperature, so at low disorder $C_v$ increases. For large disorder 
the gap vanishes, and the band expands beyond the edges, thus new excited 
states are coming out around the band-edges at the cost of the states around 
the $E_F$ (band-center) as the total number of states is fixed for the system. 
So, for large w the DOS falls around $E_F$, hence apart from a initial rise $C_v$ 
will decrease at low temperature. 

\begin{figure}[ht]
  \centering

  \begin{tabular}{cc}
   \includegraphics[width=35mm, height=28mm]{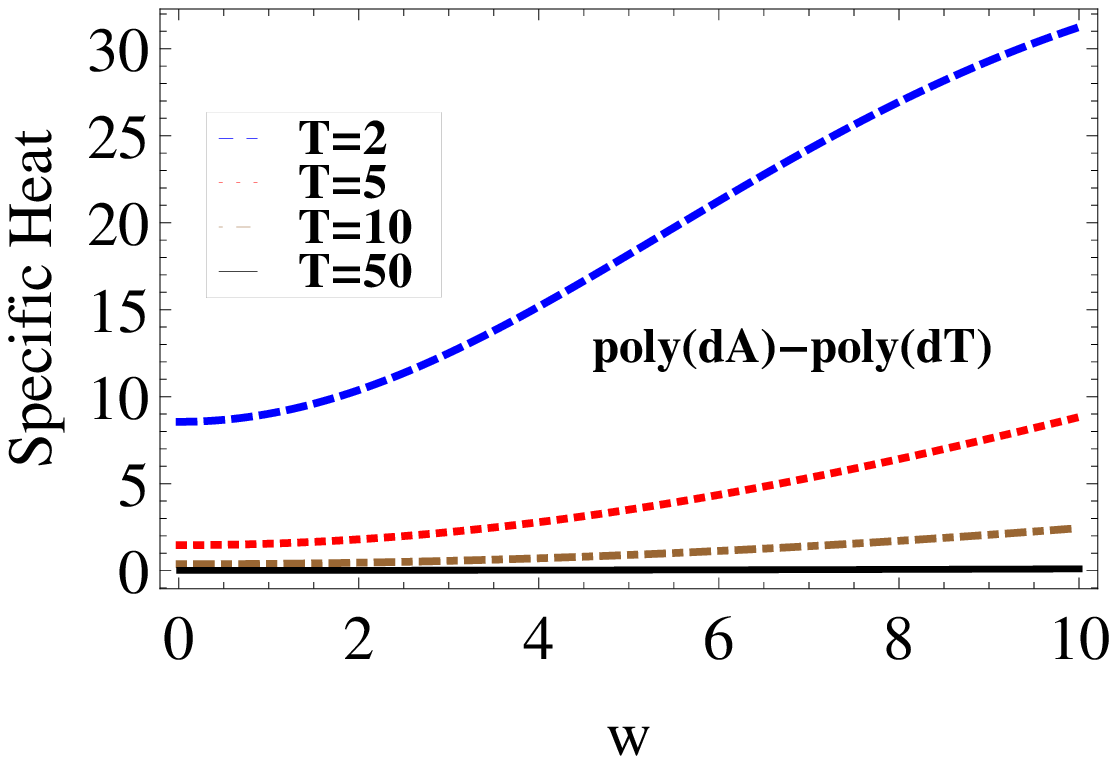}&
    \includegraphics[width=35mm, height=28mm]{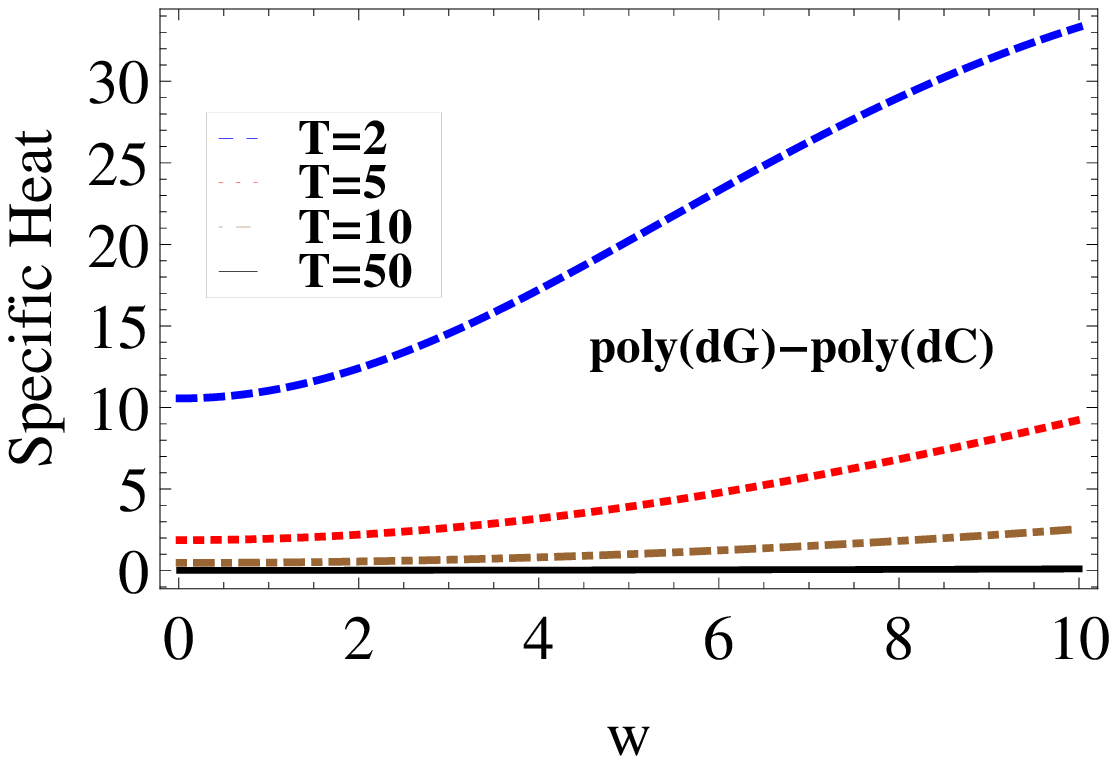}\\
    
     \includegraphics[width=35mm, height=28mm]{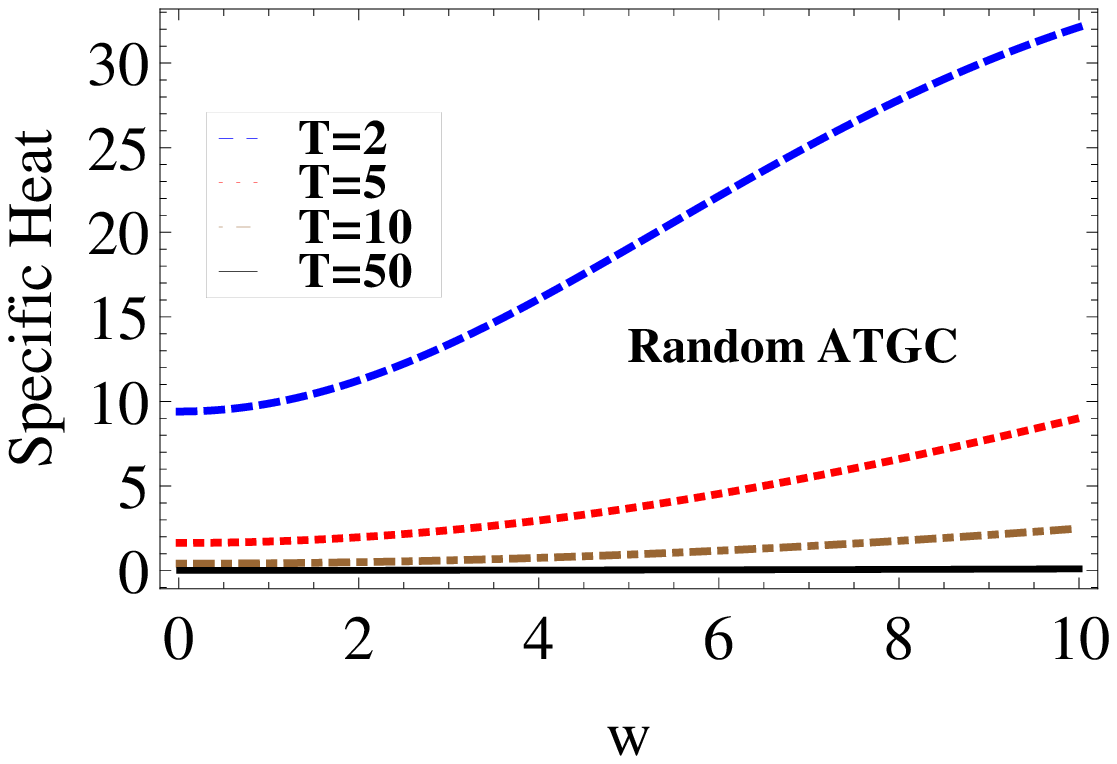}&
      \includegraphics[width=35mm, height=28mm]{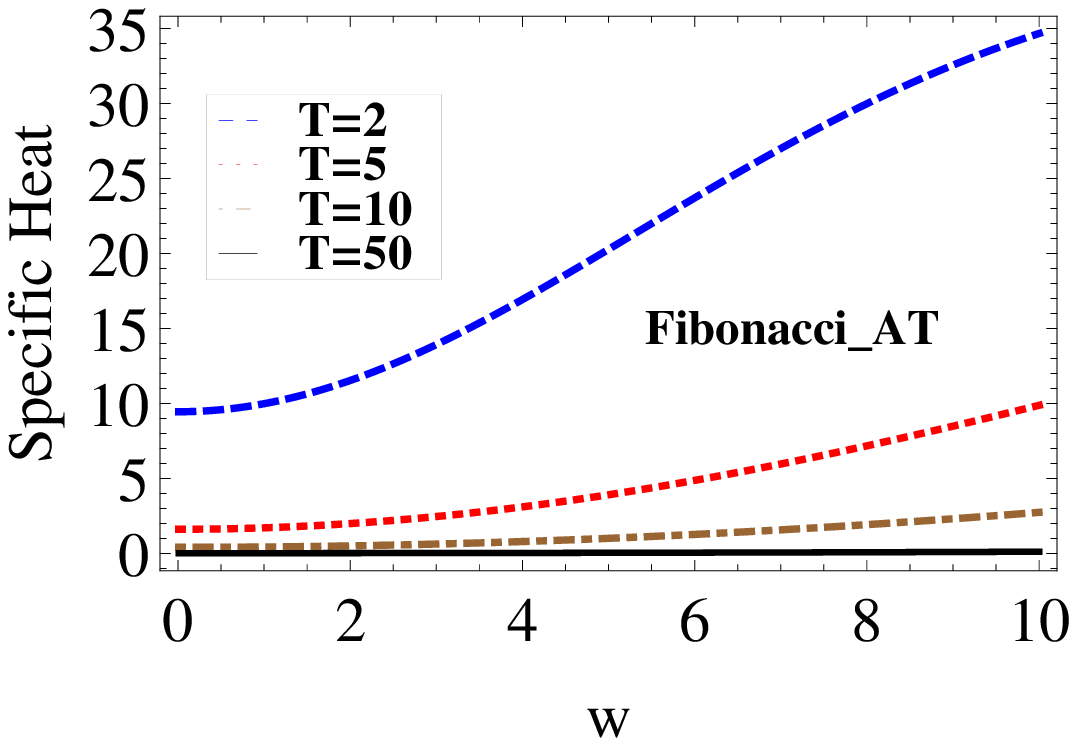}
  \end{tabular}

\caption{(Color online) Electronic specific heat $(C_v)$ as a function 
of backbone disorder strength (w) for four different DNA sequences at 
high temperature ($T>$2K). $C_v$ increases monotonically with disorder 
(w) for all the cases.}

\label{fig5}
\end{figure}

 In Fig.~\ref{fig5} we show the variation of $C_v$ with backbone disorder 
strength (w) for the high temperature (T$>$2K) range, and it exhibits that 
$C_v$ increases with temperature. To explain these we once again look at 
the corresponding DOS profile presented in Fig.~\ref{fig4}. 
At the high temperature all the states are accessible, now as we increase 
disorder, new states are appearing around the band-edges and the energy of 
these states also increase with w. As these states has high energy cost, 
the rate of change of average energy $<E>$ with T also gets increased as 
we increase w and consequently $C_v$ increases with T in the high temperature 
regime.

\begin{figure}[ht]
  \centering

 \begin{tabular}{cc}
   \includegraphics[width=35mm, height=28mm]{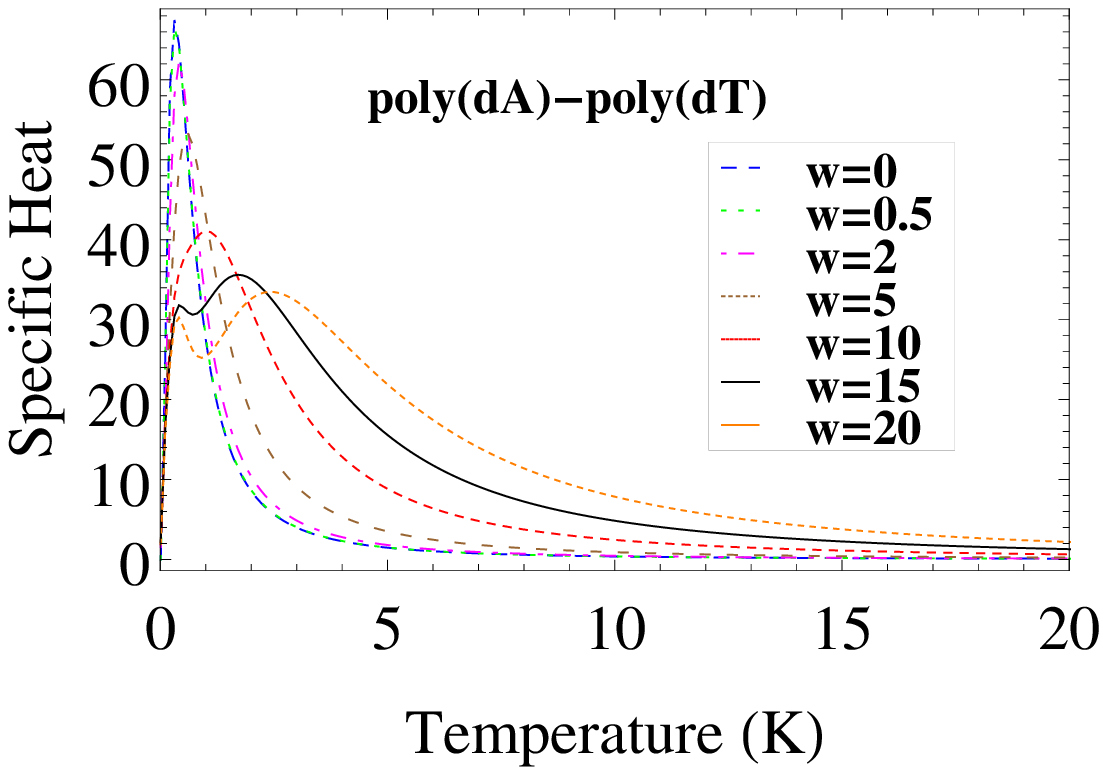}&
    \includegraphics[width=35mm, height=28mm]{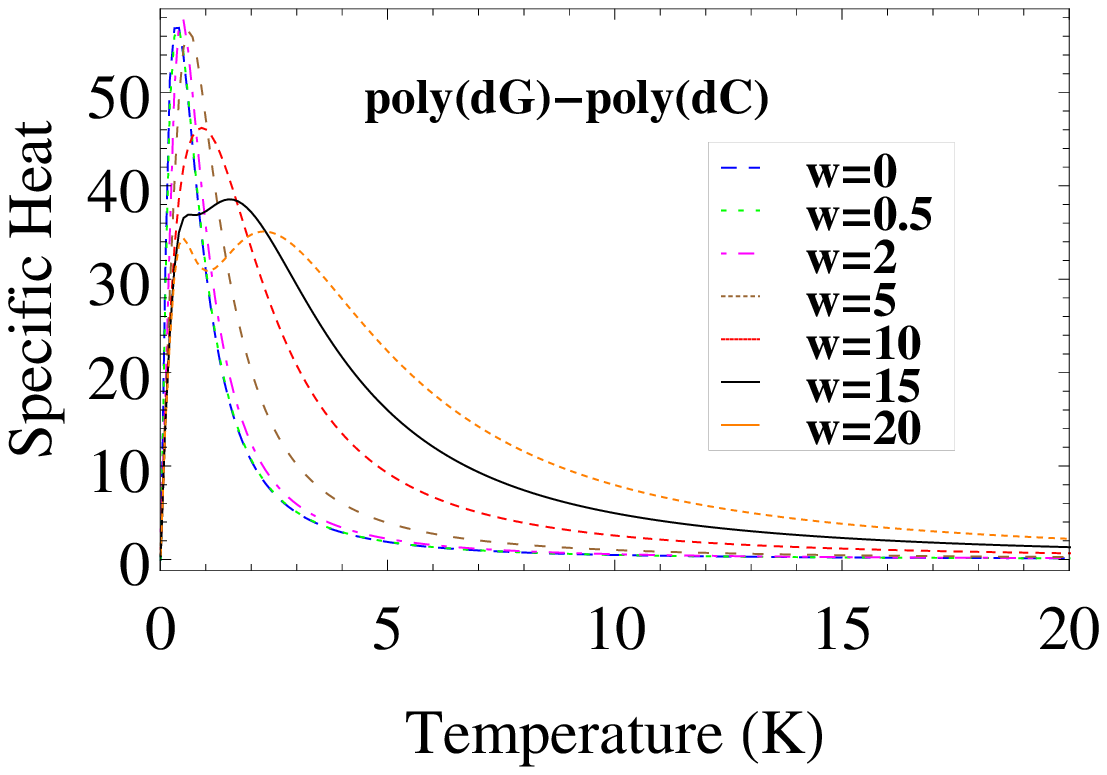}\\
     \includegraphics[width=35mm, height=28mm]{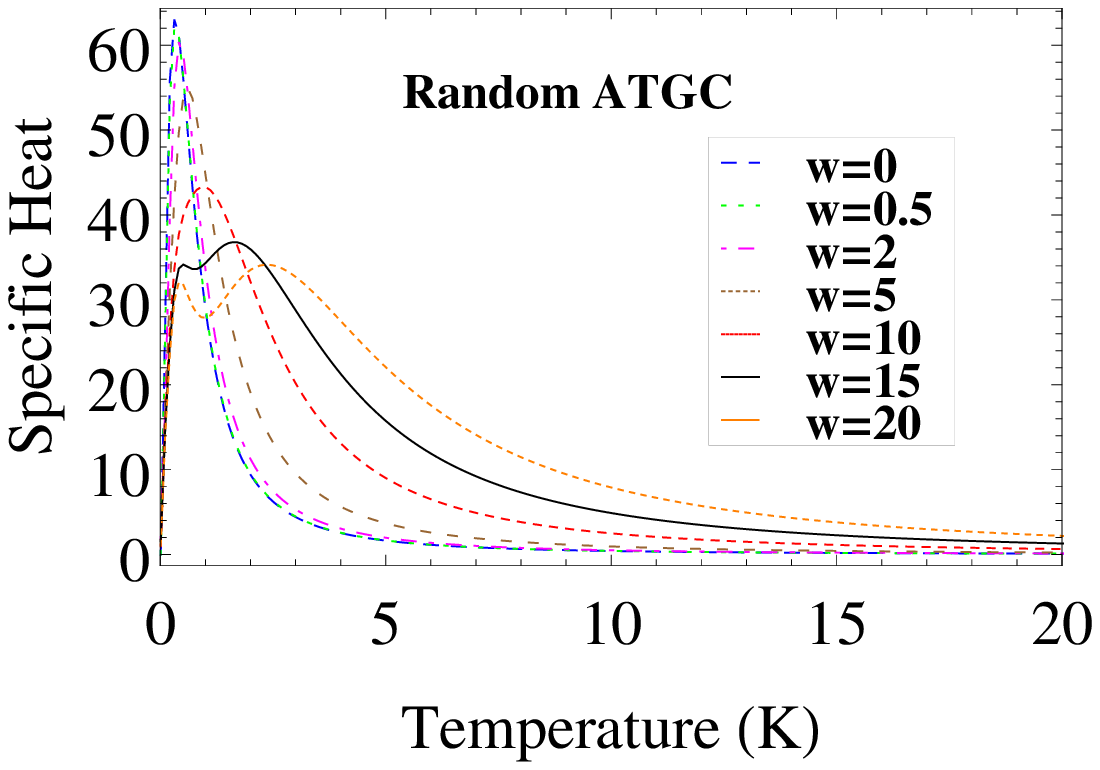}&
      \includegraphics[width=35mm, height=28mm]{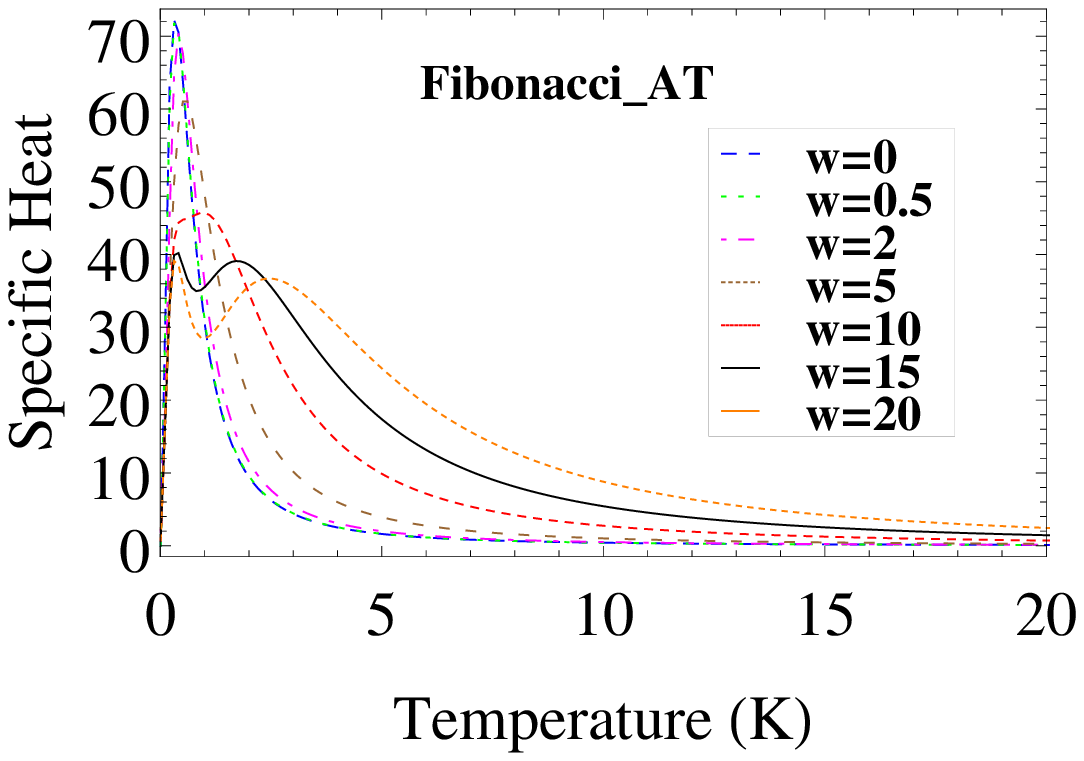}
 \end{tabular}

\caption{(Color online) Variation of electronic specific heat ($C_v$) with 
temperature ($T$) at different disorder strength (w) for four DNA sequences. 
Uniform behavior of $C_v$ under disorder irrespective of the sequences.}

\label{fig6}
\end{figure}

 In Fig.~\ref{fig6} we show the variation of $C_v$ with temperature (T) 
for four DNA sequences at various backbone disorder strength (w). In 
all the plots, at low temperature $C_v$ reduces with increasing w while 
it gets enhanced with w at high temperature as discussed earlier. The 
peak of the $C_v$ vs. T curve, which determines the crossover temperature, 
also decreases and shifts towards the high temperature as we increase 
disorder strength. There exists some fluctuations in $C_v$ at low temperature 
under sufficiently high disorder (w$>$5). Earlier this kind of oscillatory 
behavior of ESH was reported by E. L. Albuquerque {\it et. al.}~\cite{albu} 
at low temperature for the quasi-periodic sequences only, here we get the 
same kind of fluctuations for all the sequences at low temperature due to 
environmental effects.

\begin{figure}[ht]
  \centering

 \includegraphics[width=61mm]{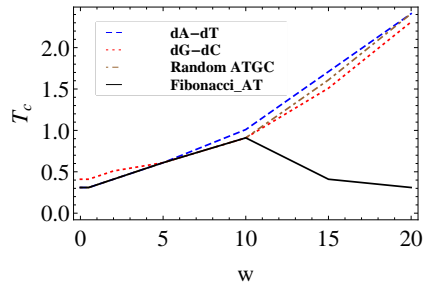}
   
\caption{(Color online) Variation of crossover temperature ($T_c$) 
with backbone disorder strength (w). Except the Fibonacci one 
the nature of variation is almost same for all the sequences.}

\label{fig7}
\end{figure}

 In Fig.~\ref{fig7} we show the dependence of crossover temperature ($T_c$) 
({\it i.e.}, the temperature at which $C_v$ becomes maximum) on backbone 
disorder (w). It shows that $T_c$ increases monotonically with w, 
the rate of increase being not uniform everywhere, for all the DNA 
sequences excepting the quasi-periodic Fibonacci one. For Fibonacci 
sequence $T_c$ increases upto a certain disorder strength and then 
it decreases. 

\section{Concluding Remarks}

 Till date thermal properties of DNA and other biomolecules are not yet 
well explored. We make an attempt to examine the electronic specific 
heat response of DNA by modelling it by the tight-binding Hamiltonian. 
Though there are some results available in the literature on DNA specific 
heat~\cite{moreira1, sarmento, moreira2, moreira3, mauriz, albu}, but they 
did not take into account the backbones being a very basic structure of 
the DNA molecules. In this work we make an attempt to study the effect 
backbone and also the effect of the environment on the electronic specific 
heat of DNA. It comes out that the introduction of backbones make drastic 
changes in the DOS profile, the band structure of the system splits up opening 
a gap in the central region of the band. Due to the formation of this gap specific 
heat gets enhanced in presence of backbones over the entire temperature range 
excepting a narrow low temperature region. On environmental fluctuations, $C_v$ exhibits 
two distinct behaviours, at low temperature it decreases with backbone disorder 
strength (w) and in the high temperature region it increases with w. We have also 
seen that the cross-over temperature ($T_c$) which corresponds to the maximum 
of the specific heat vs temperature curve increases with disorder (w). In 
this way we have been able to put forward a regularized behavior of the ESH 
of DNA being independent of the sequence we have chosen. The effect of 
environmental fluctuations on ESH is quite universal both for the clean 
case (w=0) and also in presence of environmental disorder. It implies that 
ESH of the system reacts to the environment in the same way irrespective 
of its sequential variety. In order to verify our predictions experimentally, 
heat exchange in the process of protein binding, unfolding, ligand association 
and other bimolecular reactions should be measured with much more reliability. 
There are three basic techniques used for this measurements, the differential 
scanning calorimetry (DSC) which measures sample heat capacity with respect 
to a reference as a function of temperature, isothermal titration calorimetry 
(ITC) which measures the heat absorbed or rejected during a titration experiment 
and the third one is thermodynamic calorimetry (for a detail description see 
Ref.~\cite{jelesarov}). But unfortunately none of these techniques is able 
to separate the electronic contribution to the specific heat of biomolecules, 
as they measure all contributions including the vibrational ones. However, 
we hope that there will be experimental verification of our results and 
other investigations to find thermal properties of DNA and alike biomolecules 
in near future with modifications of the above mentioned tools. 

\section{Appendix} 

If we take the temperature dependence of chemical potential explicitly 
into account then the expression of specific heat becomes 

\begin{eqnarray}
C_v = \frac{1}{4 k_B T^2} [\sum\limits_{i=1}^N E_i^2 cosh^{-2}[\frac{(E_i-\mu)}{2k_BT}]-\nonumber\\
\frac{(\sum\limits_{i=1}^N {E_i cosh^{-2}[\frac{(E_i-\mu)}{2k_BT}]})^2}{\sum\limits_{i=1}^N cosh^{-2}[\frac{(E_i-\mu)}{2k_BT}]}]
\end{eqnarray} 

where the chemical potential ($\mu=\mu(n_e/N,T)$) can be obtained numerically 
form the Fermi distribution following 

\begin{equation}
n_e= \sum\limits_{i=1}^N <f(E_i)>
\end{equation} 
Here, $n_e$ is the number of non-interacting electrons and N is
the total number of one-particle accessible states in the system. 
We check our results using Eq.9, but find no significant changes.

\end{document}